\documentclass[traditabstract]{aa}
\usepackage{natbib}
\usepackage{graphicx}
\usepackage{epsfig}
\usepackage{txfonts}

\def\msol{M$_\odot$}             %Msun
\def\mjup{M$_{Jup}$}             %Msun
\def\a{HD\,1690\,}
\def\b{BD\,-114672\,}
\def\c{HD\,25171\,}
\def\d{HD\,33473A\,}
\def\e{HD\,72659\,}
\def\f{HD\,89839\,}
\def\g{HD\,113538\,}
\def\h{HD\,167677\,}
\def\i{HD\,217786\,}
\def\j{HIP\,21934\,}

\begin{document}

\title{ The HARPS search for southern extra-solar planets\thanks{Based
on observations made with the HARPS instrument on the ESO 3.6 m
telescope at La Silla Observatory under programme IDs 072.C-0488(E) and 085.C-0019.}
}

\subtitle{XXVI: Seven new planetary systems }

\author{
C. Moutou\inst{1}
\and M. Mayor\inst{2}
\and G. Lo Curto\inst{3}
\and D. S\'egransan\inst{2}
\and S. Udry\inst{2}
\and F. Bouchy\inst{4,5}
\and W. Benz\inst{6}
\and C. Lovis\inst{2}
\and D. Naef\inst{2}
\and F. Pepe\inst{2}
\and D. Queloz\inst{2}
\and N.C. Santos\inst{7}
\and S. G. Sousa\inst{7}
}

\offprints{C. Moutou}

\institute{
Laboratoire d'Astrophysique de Marseille, OAMP, Universit\'e Aix-Marseille \&
CNRS, 38 rue Fr\'ed\'eric Joliot-Curie, 13388 Marseille cedex 13, France\\
\email{Claire.Moutou@oamp.fr}
\and 
Observatoire de Gen\`eve, Universit\'e de Gen\`eve, 51 ch.des Maillettes, 1290 Sauverny, Switzerland.
\and
ESO, Karl-Schwarzschild Strasse, 2, Garching bei M\"unchen, Germany
\and
Institut d'Astrophysique de Paris, 98bis bd Arago, 75014 Paris, France
\and
Observatoire de Haute Provence, OAMP, CNRS, F-06670 Saint Michel l'Observatoire, France
\and
Physikalisches Institut Universit\"at Bern, Sidlerstrasse 5, 3012
Bern, Switzerland
\and
Centro de Astrofisica e Departamento de F'sica e Astronomia, Universidade do Porto, Rua das Estrelas, 4150-762 Porto, Portugal
}

\date{Received ; accepted }

\abstract{We are conducting a planet search survey with HARPS since seven years. The volume-limited stellar sample includes all F2 to M0 main-sequence stars within 57.5 pc, where extrasolar planetary signatures are systematically searched for with the radial-velocity technics. In this paper, we report the discovery of new substellar companions of seven  main-sequence stars and one giant star, detected through multiple  Doppler measurements with the instrument HARPS installed on the ESO  3.6m telescope, La Silla, Chile.  These extrasolar planets  orbit the  stars \a, \c, \d, \f, \g, \h, and \i. The already-published giant planet around \e is also analysed here, and its elements are better determined by the addition of HARPS and Keck data. The other discoveries are giant planets in distant orbits, ranging from 0.3 to 29 \mjup\, in mass and between 0.7 and 10 years in orbital period. The low metallicity of most of these new planet-hosting stars reinforces the current trend for long-distance planets around metal-poor stars.

 Long-term radial-velocity surveys allow probing the outskirts of extrasolar planetary systems, although confidence in the solution may be low until more than one orbital period is fully covered by the observations. For many systems discussed in this paper, longer baselines are necessary to refine the radial-velocity fit and derive planetary parameters. The radial-velocity time series of stars \b and \j are also analysed and their behaviour interpreted in terms of the activity cycle of the star, rather than long-period planetary companions.
}

\maketitle

\section{Introduction}

The complementarity between planet search techniques will allow us in a mid-future to draw the general picture of extrasolar systems. The transit method is strongly biased towards short-period planets, while direct imaging will be limited to exploring the outer regions of close-by systems. Intermediately, the radial-velocity surveys, conducted for more than 15 years, are continuously discovering planets in the wide period range between 1 day and 15 years. In some more years, we will be able to study the same extrasolar systems by means of the gravitational perturbation inferred by the planets on their star (by the radial-velocity and/or the astrometric technics), and direct imaging with the aid of the next-generation high-contrast instruments. This will be possible only for a few nearby systems that contain outer giant planets, or for young systems when the planet has a higher self-luminosity than a planet of similar mass in an evolved system. The combination of both methods will permit calibrating the atmospheric models, i.e., the relationship between the mass of the planet and its spectral energy distribution (e.g., \citet{baraffe2003,fortney2008}).

In this context, the volume-limited sample conducted with HARPS  since 2003 is one of the main sources of favourable targets for future direct imaging surveys. With a precision of about 2 m/s over 850 stars, it has allowed the discovery of 32 planets so far \citep{pepe04,moutou05,locurto06,naef07,moutou09,locurto10,segransan10,naef10}. In this paper, we present the HARPS data collected for eleven stars in this stellar sample, which gathers F2 to M0 main-sequence stars closer than 57.5 pc from the Sun.  Their radial-velocity variations is analysed and interpreted as the signature of planetary companions, except in two cases where we show strong evidence for the detection of the stellar long-term magnetic cycle.
In section \ref{star}, the parent stars are described, while radial-velocity observations and their analysis are presented in section \ref{planet}; a discussion follows in section \ref{ccl}.

\section{Characteristics of the host stars}
\label{star}
The spectroscopic analysis of the stars has been conducted on the combined HARPS spectra, following the method described in \citet{sousa08}.  The method is based on the equivalent widths of Fe I and Fe II weak lines, by imposing excitation and ionization equilibrium assuming LTE. For this task the 2002 version of the code MOOG  was used \citep{sneden73} and a grid of Kurucz Atlas 9 plane-parallel model atmospheres \citep{kurucz93}. Equivalent widths of spectral lines are estimated with the ARES code  \citep{sousa07}, from which the values of T$_{eff}$, log$g$ and [Fe/H] are derived. Actual errors on these parameters are obtained by quadratically adding 60 K, 0.1 and 0.04 dex respectively to the internal errors on T$_{eff}$, log$g$ and [Fe/H]. These values were estimated by considering the typical dispersion in each parameter comparison plot with the infra-red flux method \citep{casagrande06}, presented in Figures 3 to 5 of \citet{sousa08}. The stellar luminosity was calculated based on Hipparcos photometry \citep{hip} and revised Hipparcos parallaxes \citep{vanleeuwen07}, including the bolometric correction as calibrated in \citet{flower96}. Theoretical evolutionary tracks are compared with the measured values to estimate the mass, radius and age of the stars with a Bayesian analysis\footnote{see http://stev.oapd.inaf.it/} \citep{dasilva06}. The width and contrast of the cross-correlation function are measured to estimate the projected rotational velocity of the star. Finally, the activity indicator log$R'_{HK}$ is measured on individual spectra when the flux is sufficient at 390 nm \citep{santos02}.

Tables \ref{stars1} and \ref{stars2} summarize the measured and derived parameters of the stars. For the reddest stars, the relationships between the cross-correlation function and the stellar parameters (particularly vsin$i$) are not valid \citep{santos00,santos02}, and those values are thus missing.

\begin{figure}[h]
\epsfig{file=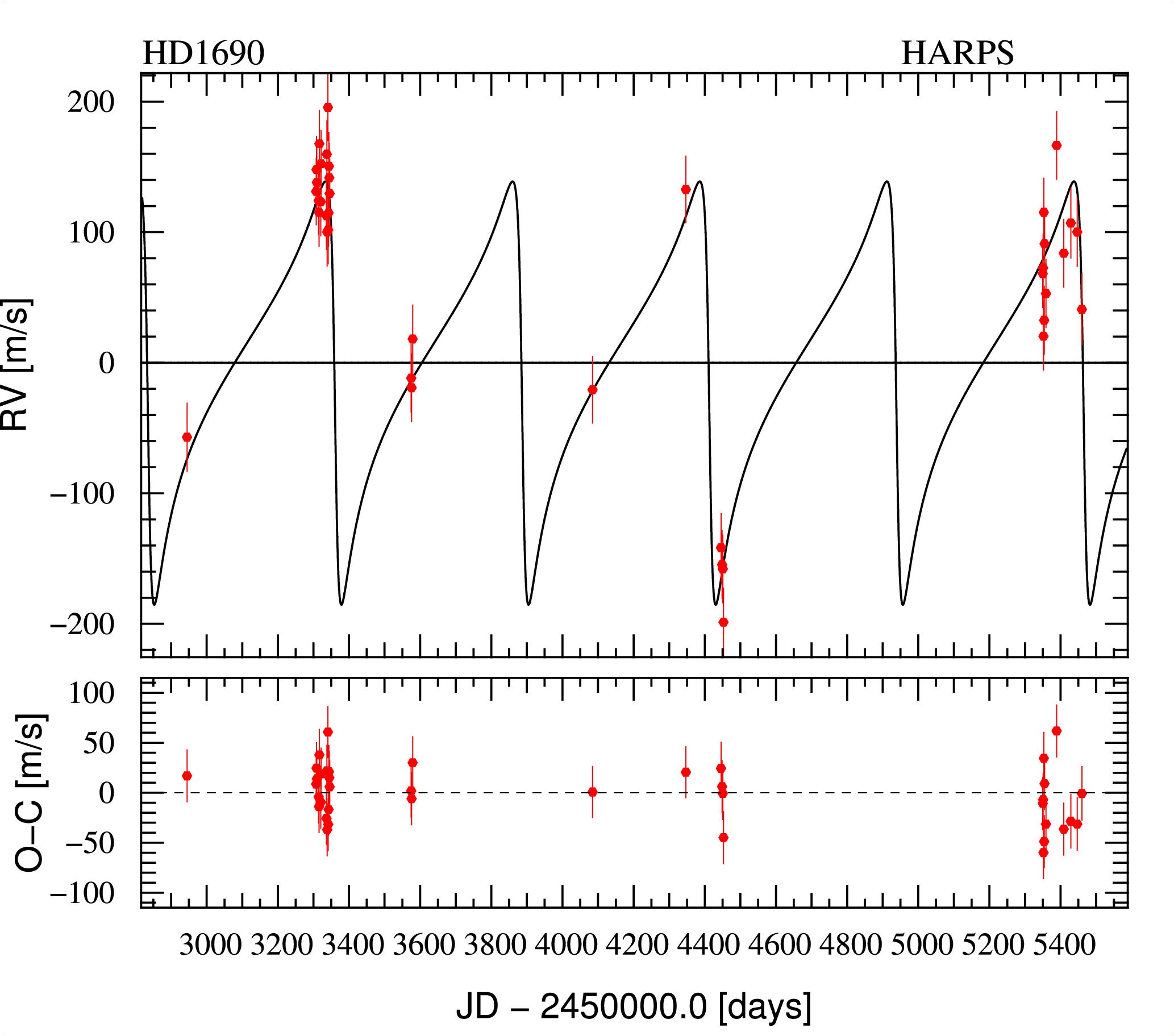,width=8cm}
\epsfig{file=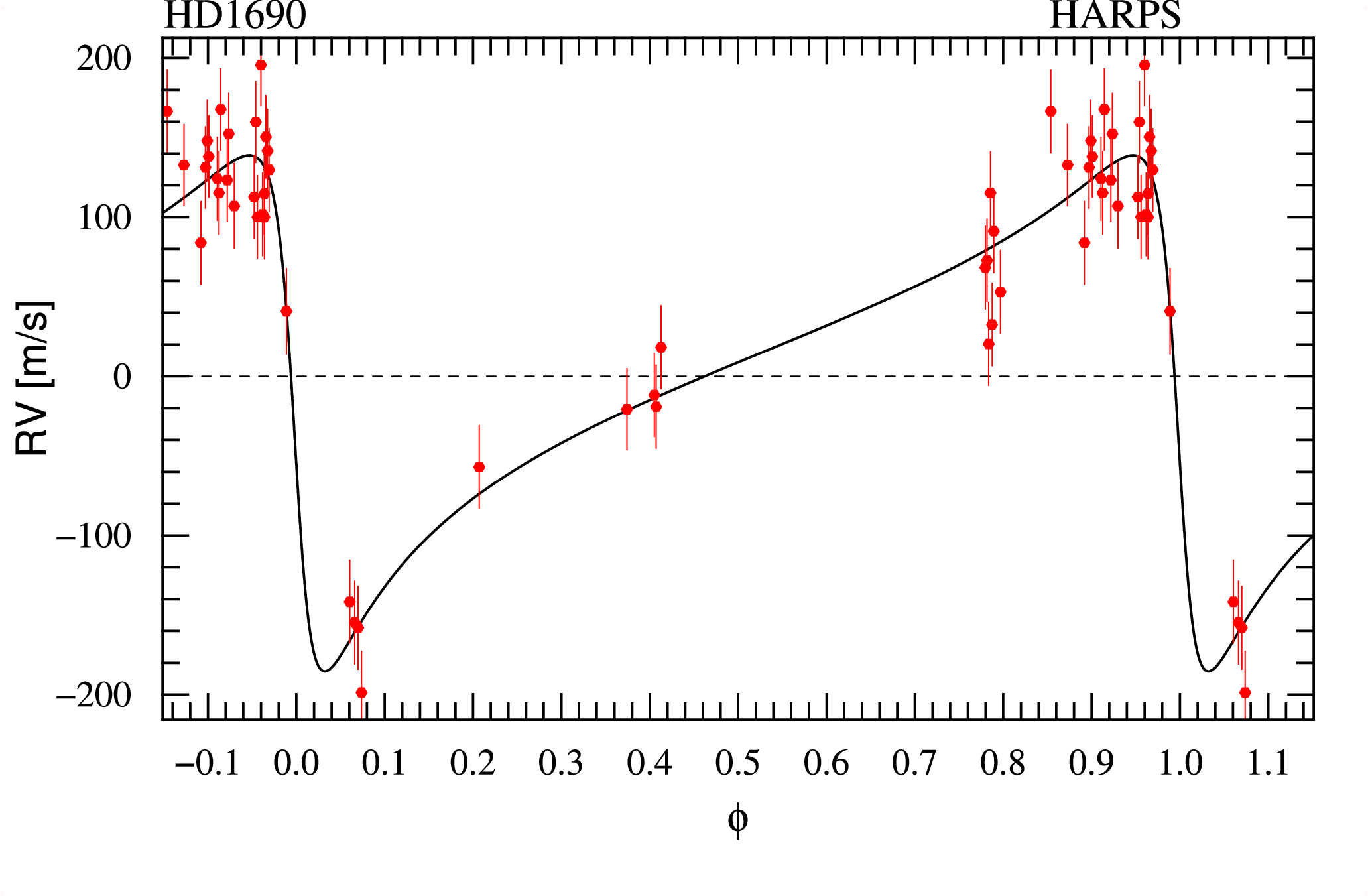,width=8cm}
\caption{Left: Radial-velocity measurements of \a obtained with HARPS against time. The orbital fit  is overplotted as a black line. The residuals to the fit are given below. Right: the phase-folded radial-velocity curve is shown. The best fit to the data is shown as a plain line.}
\label{timea}
\end{figure}

\begin{figure}[h]
\epsfig{file=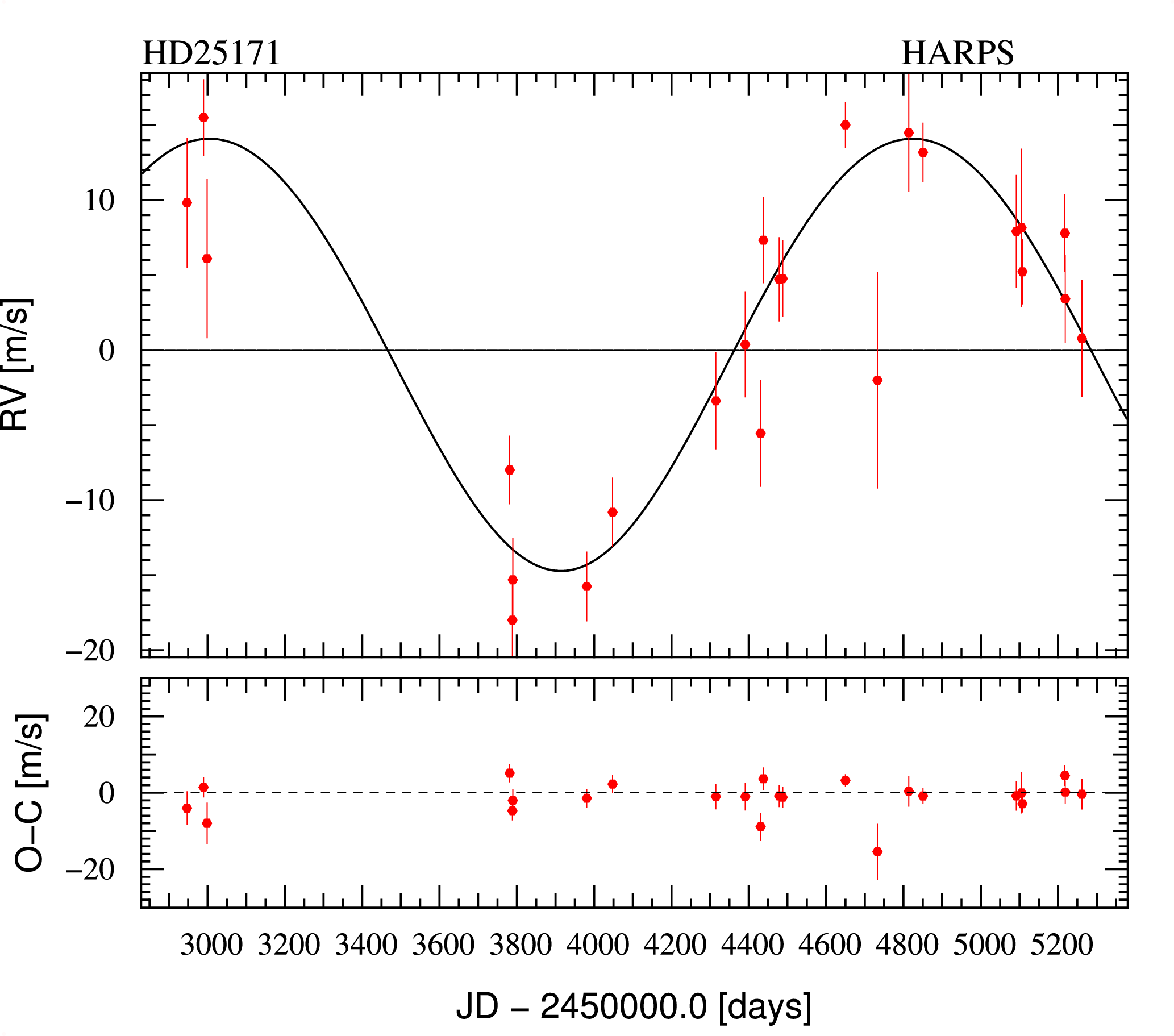,width=8cm}
\caption{Radial-velocity measurements of \c obtained with HARPS against time and the residuals to the best-fit model. }
\label{timec}
\end{figure}
\begin{figure}[h]
\epsfig{file=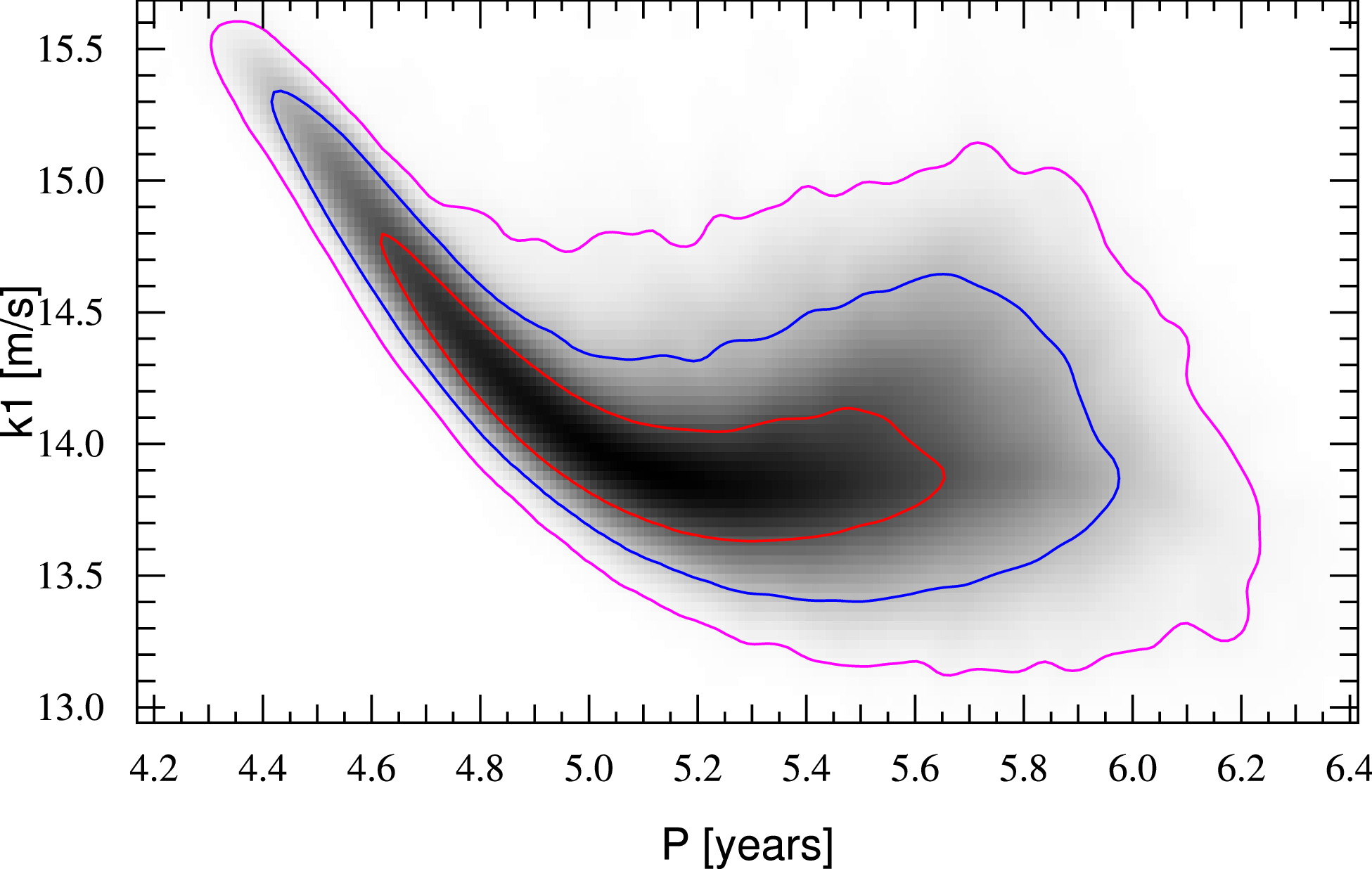,width=8cm}
\epsfig{file=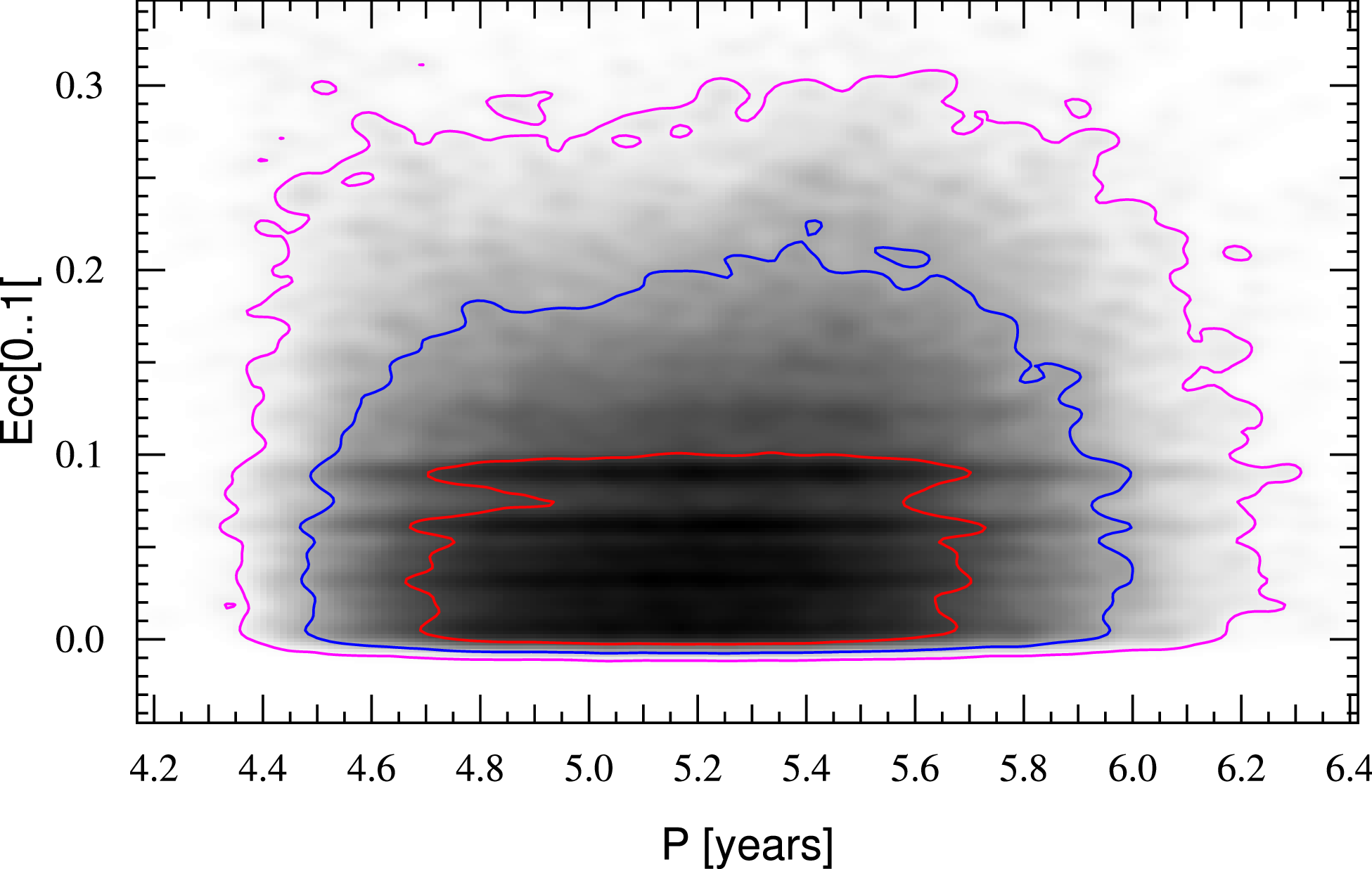,width=8cm}
\caption{Variations of $\chi^2$ as a function of various related parameters of the fit for HD 25171, using Markov-chain simulations. (Left)  Correlation plot between the orbital period and radial-velocity semi-amplitude. (Right) Correlation plot between eccentricity and orbital period.  The lines show 1,2 and 3-$\sigma$ iso-contour plots.  }
\label{statc}
\end{figure}

\section{Radial velocity data and proposed solutions}
\label{planet}

Radial-velocity time series of our sample stars have been obtained with the spectrograph HARPS \citep{messenger03}, mounted at the 3.60m telescope in La Silla, ESO, Chile, mostly in the framework of the Guaranteed Time Observation programme (programme ID 072.C-0488), with additions from the follow-up volume-limited search programme 085.C-0019. 

We used the HAM mode with a spectral resolving power of 115000, without simultaneous calibration of the velocity by the ThAr calibration lamp: the spectrograph is indeed stable enough for the precision of about 2 m/s  targeted in this survey. Such a precision corresponds to a signal-to-noise ratio of 40 at 550 nm and to exposure times ranging from 2 to 15 min according to the stellar magnitude and external conditions, with a mean value of 8 minutes.

Each individual spectrum was then cross-correlated with a numerical mask, built from the stellar template better matching the stellar type, from M0 to F0 \citep{baranne1996}. In addition to the position of the cross-correlation peak,  the slope of the bisector line is measured on the stellar profile, indicating the level of photospheric activity that may affect the measurement of the peak position. The variation of the bisector span with the radial velocity is shown in Figure \ref{allbis} for all stars of the presented sample. Let us recall that an anti-correlation of the bisector span with respect to the peak position may depict a spot-related variation of the radial velocity, especially when the amplitudes of both variations are similar. Bisector variations may also be used to diagnose long term radial-velocity signals induced
by variations in the stellar magnetic cycle \citep{santos2010a}, as well as variations in the width of the cross-correlation function and temporal evolution of the activity tracer log$R'_{HK}$ (see section 3.3) .

In the time series of stellar radial velocities presented below, a significant variation was detected over periods usually longer than several hundreds days. After checking for activity indicators, we present an analysis of the results in terms of the planetary orbital signature except for two cases where the long-term activity cycle seems the dominant factor. In a few cases, several solutions are equivalent and more data over monthes to years are needed to conclude. Radial-velocity time series and the proposed solution are presented in Figures \ref{timea} to \ref{timecy}. Individual measurements are available in electronic form at the CDS via anonymous ftp to cdsarc.u-strasbg.fr (130.79.128.5) or via http://cdsweb.u-strasbg.fr/cgi-bin/qcat?J/A+A/. Error bars for the derived parameters were estimated from a Monte-Carlo bootstrap analysis of the data sets using the routines of {\sc yorbit} (written by one of us, D.S.). 

\subsection{Complete orbits}

\subsubsection{\a}
 Fourty-two HARPS measurements of the giant star  \a were gathered from November 2003 to October 2010, as shown in Figure \ref{timea}.  The presence of a few giant stars in the HARPS volume-limited sample is due to a revision of the parallax from the original Hipparcos output catalogue \citep{hip} to the new estimates of parallaxes from \citet{vanleeuwen07}. For star \a, the parallax was revised from 43.42 to 3.22 mas.

The average formal accuracy of individual data points is 1.8 m/s. Variations of the radial velocity have a standard deviation of 110 m/s, while the bisector slope varies with a standard deviation of less than 20 m/s.

When the variation of the stellar radial velocity of \a is interpreted in terms of a Keplerian orbit of a planet, the best solution corresponds to a  532 day period.  The residual scatter of the fit is $rms=$ 25.3 m/s. Since the parent star is a giant, one expects the residual scatter to originate from high-frequency stellar oscillations, with an amplitude that is standard for red giants \citep{sato2005}. No significant period is found in the residual of the fit, which is also compatible with the common behaviour of radial-velocity oscillation signatures in giants \citep{hatzes2007}. In order to better reflect the actual velocity jitter and to get a reduced $\chi^2$ closer to 1, we quadratically added 25 m/s to the individual errors.  This jitter amplitude is common for a K giant with $B-V$=1.35, a range where it may vary from 20 to 100 m/s  \citep{frink01}.  The residuals show a slight negative slope, indicating a possible distant companion. The solution implying a linear drift reproduces the data better than the single planet with a probability of 75\% (F-test of \citet{pourbaix01}). The orbit of the planet shows  large eccentricity, $e=$ 0.64, leading to a minimum mass of the companion of about 6 \mjup. Figure \ref{timea} shows the phase-folded data points and best-fit orbit.
 It must be noted that the solution shown in this analysis is not very robust: the removal of $\sim$12 data points significantly changes the period of the best-fit. The reason for this are several peaks of equivalent power in the periodogram. The presented solution is found by performing 10000 Monte Carlo simulations and including all data points.  The complete solution is given in Table \ref{sol1}. It is possible that further observations modify the solution proposed here, or increase the number of planets in the system.
 
\subsubsection{\c}
The non-active star \c was observed 24 times with HARPS between November 2003 and March 2010 with average errors of 3 m/s.  A long-period variation is detected and attributed to the presence of a distant Jupiter-like planet. The standard deviation of the velocity drops by a factor 3 when a best-fit Keplerian orbit is removed from the data.  A period of 5 years is found, corresponding to a planet minimum mass of 0.95 \mjup\,  (Figure \ref{timec}). Since the semi-amplitude is low and the long orbit only partially covered by observations, we performed a Monte-Carlo Markov-chain analysis to support the detection. The most relevant output plots are shown on Figure \ref{statc}: the $\chi^2$ distribution of the fit, against related parameters such as the semi-amplitude, the eccentricity, and the orbital period. The correlation plots show that several equivalent solutions exist, but in a limited range of these parameters
The complete solution is given in Table \ref{sol1} from the bootstrap analysis that gives more conservative errors than the MCMC analysis.
The scatter of the residuals to this fit is 3.4 m/s, which is of the order of individual measurements.

\subsubsection{\e}
\begin{figure}[h]
\epsfig{file=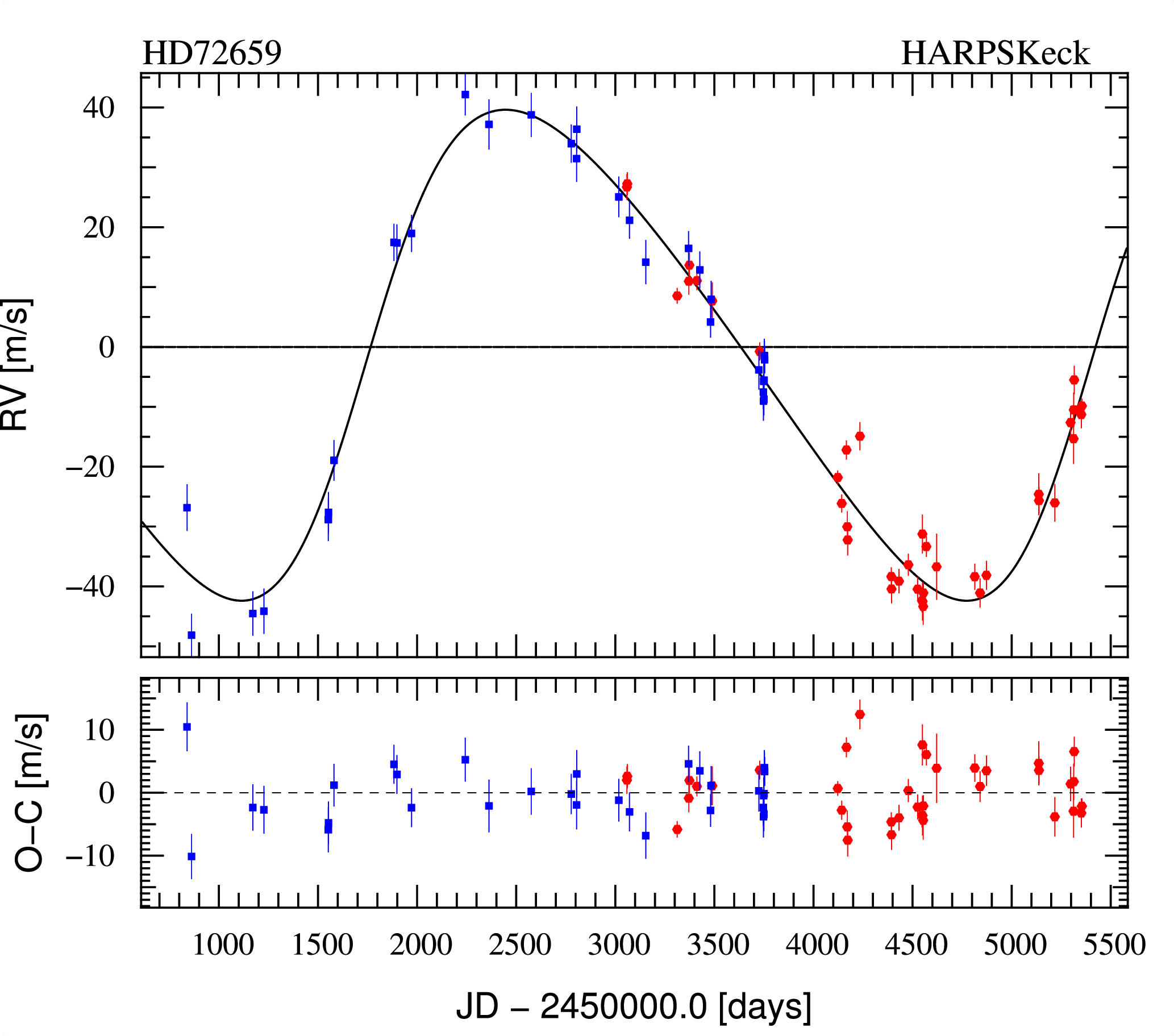,width=8cm}
\caption{Radial-velocity measurements of \e obtained with HARPS (red points) and Keck (blue points) against time, together with the best-fit solution of the combined data sets and the residuals. }
\label{timee}
\end{figure}

A total of 36 spectra of \e have been gathered between February 2004 and June 2010, with an average error of 2.3 m/s (Figure \ref{timee}, red points). The radial-velocity time series shows variations with a maximum amplitude of 70 m/s, whereas the bisector slope is stable to 6 m/s. The full period is seemingly not covered by our observations. The low activity of the star and the absence of a correlation of the bisector span or activity tracers with the radial velocity exclude the possibility that activity alone explains the velocity variation. 

A planetary companion has already been reported by \citet{butler2003} in this system, with poor constraint on the orbit. A refined solution is published in \citet{butler2006}. The combination of both Keck/HIRES and HARPS data reinforces the solution with more than 12 year span and in particular, the orbital period is now well determined. The offset between the Keck and HARPS data sets is left as a free parameter to the fit.
Our solution is compatible with the one given in \citet{butler2006}, but with an error on the period seven times smaller. The best-fit solution shows the presence of a single orbiting 3.1-\mjup\, planet in a slightly eccentric orbit of period 3658 days (Figure \ref{timee}). The residuals of the fit have a standard deviation of 3.7 m/s. No long-term trend or significant peak in the periodogram of residuals is found to explain this low-amplitude extra jitter. The complete solution is given in Table \ref{sol1}.

\subsubsection{\g}
\begin{figure}[h]
\epsfig{file=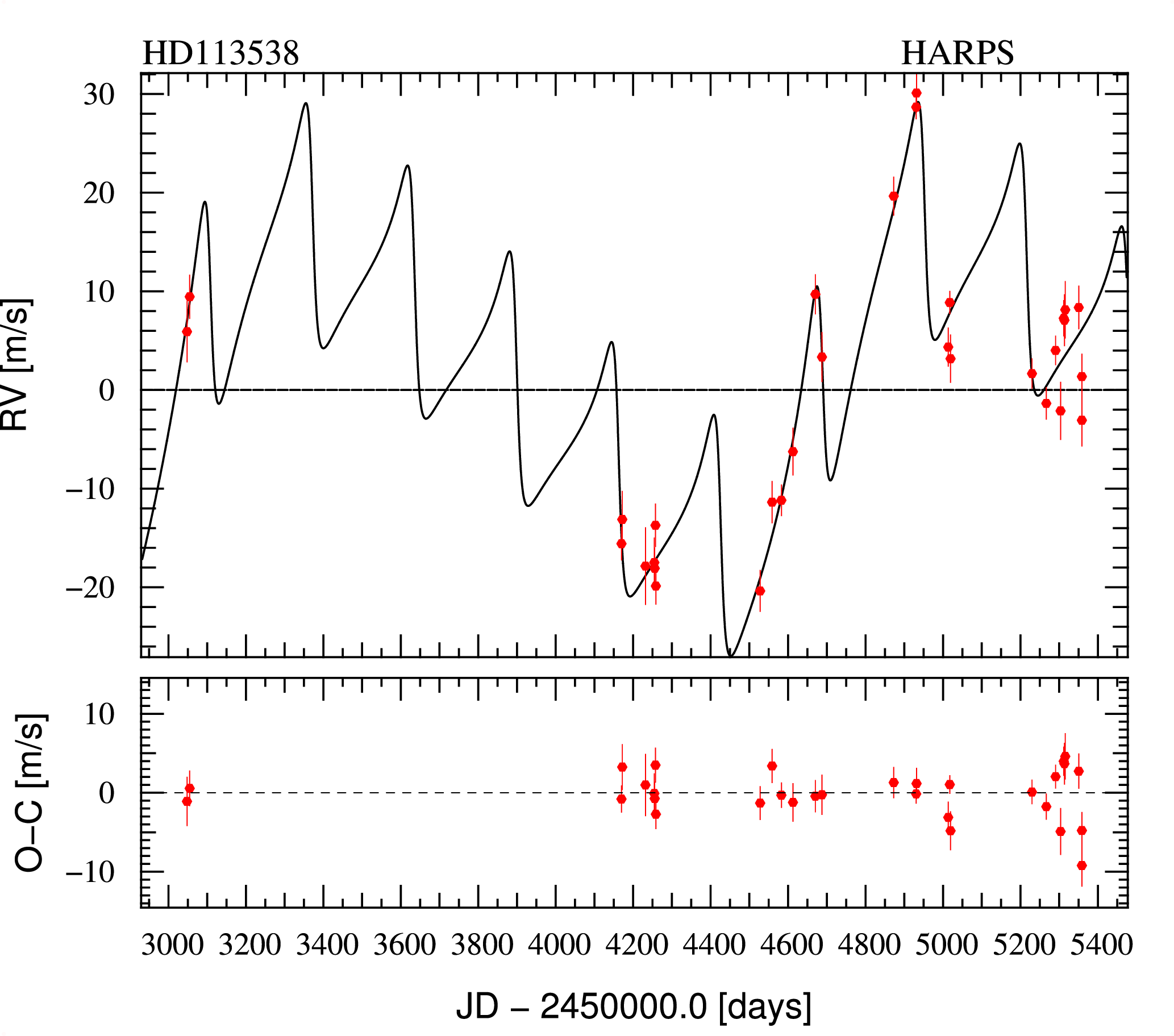,width=8cm}
\epsfig{file=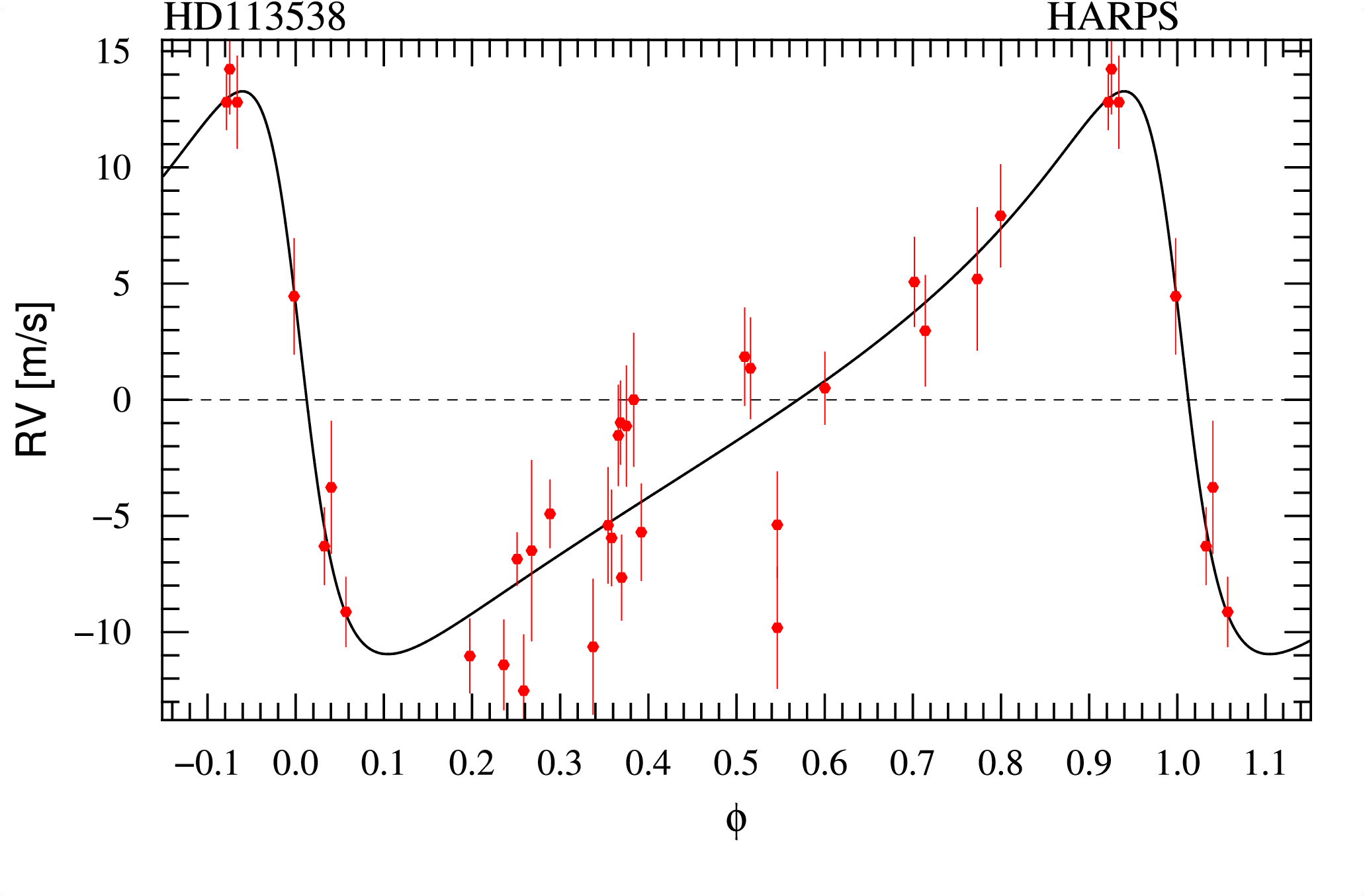,width=8cm}
\epsfig{file=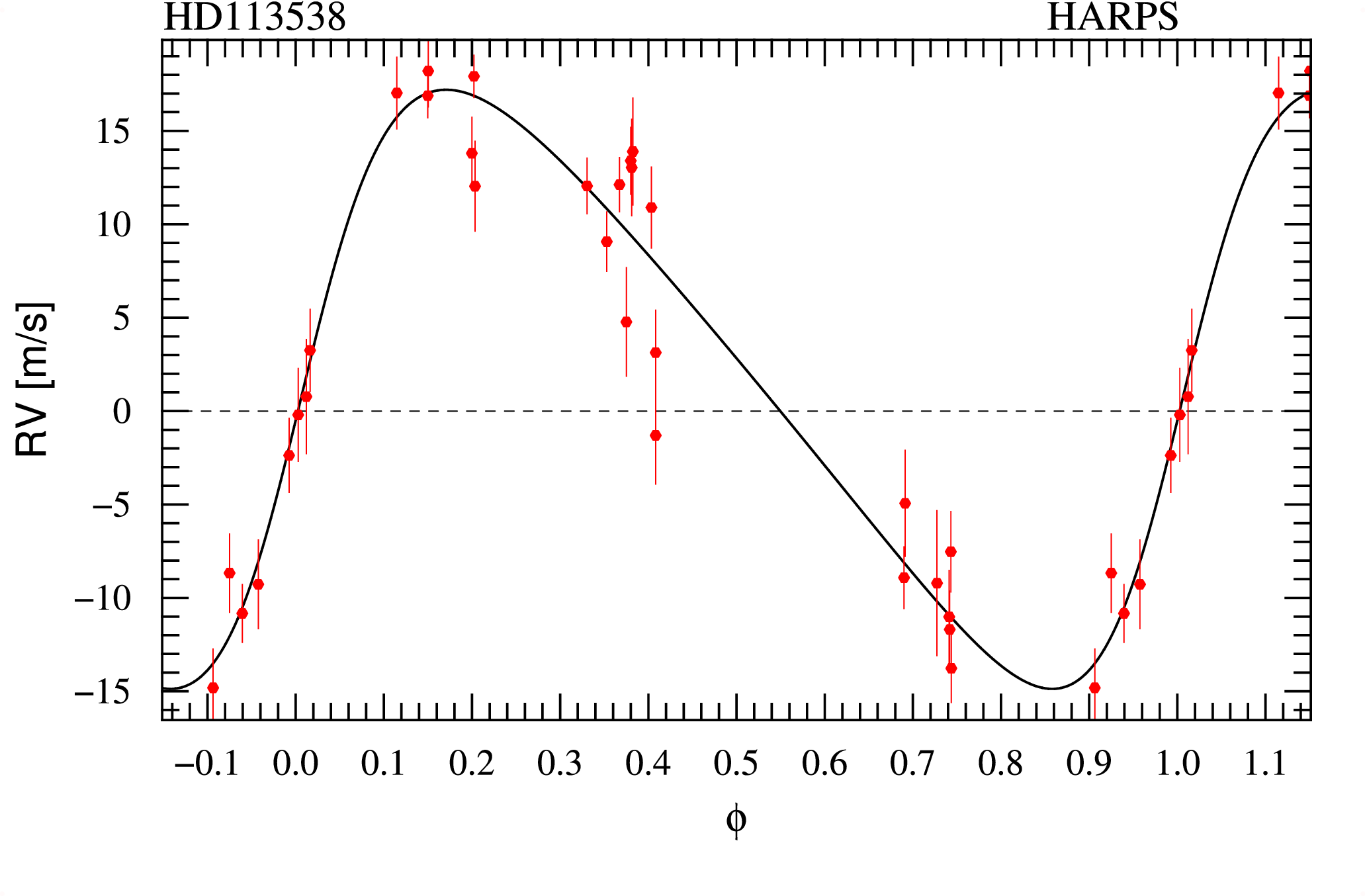,width=8cm}
\caption{Radial-velocity measurements of \g obtained with HARPS against time, and for both planets, against their orbital phase. }
\label{timeg}
\end{figure}

We secured 29 HARPS measurements of star \g between February 2004 and June 2010; they have average errors of 1.8 m/s. The radial velocity variations have a peak-to-peak amplitude of 50 m/s over the 2268 day time span of the observations, and they do not correlate with the bisector span variations. The best-fit solution involves a single planet with a long period and very eccentric orbit, but residuals to the fit are large with regard to individual errors (standard deviation of 5.4 m/s), and with a poorly defined semi-amplitude. A better solution is depicted with a two-Keplerian orbit model, involving a Jupiter-like planet in a slightly eccentric orbit of 1657 day period, plus a Saturn-like planet in a shorter eccentric orbit of 263 days. Then the residuals to the combined fit drop to 1.8 m/s. The complete solution is given in Table \ref{solg} and Figure \ref{timeg}. Simulations were performed to confirm that the two-planet solution is statistically more significant than the single planet $+$ long-term trend variation:  the data are permuted around the one-planet solution, and the residuals are statistically compared with the two-planet model residuals. The probability that the two-planet model suits the data better is obtained by calculating the number of realisations where the $\chi^2$ obtained with the permuted data is smaller than with real data. The use of the genetic algorithm in this iterative process allows more reliability than false alarm probability tests on the periodogram. 

 As an additionnal test and because the star is significantly active, we used the activity indicators to check whether the signal is caused by activity. The FWHM of the cross-correlation function shows a moderate correlation to the Mount Wilson $S$ activity index, which is based on calcium emission, with a correlation coefficient of 60\%. Our previous analysis shows that the two-planet model is statistically robust, but we remain cautious because stellar activity probably affects the solution.

\subsubsection{\i}
\begin{figure}[h]
\epsfig{file=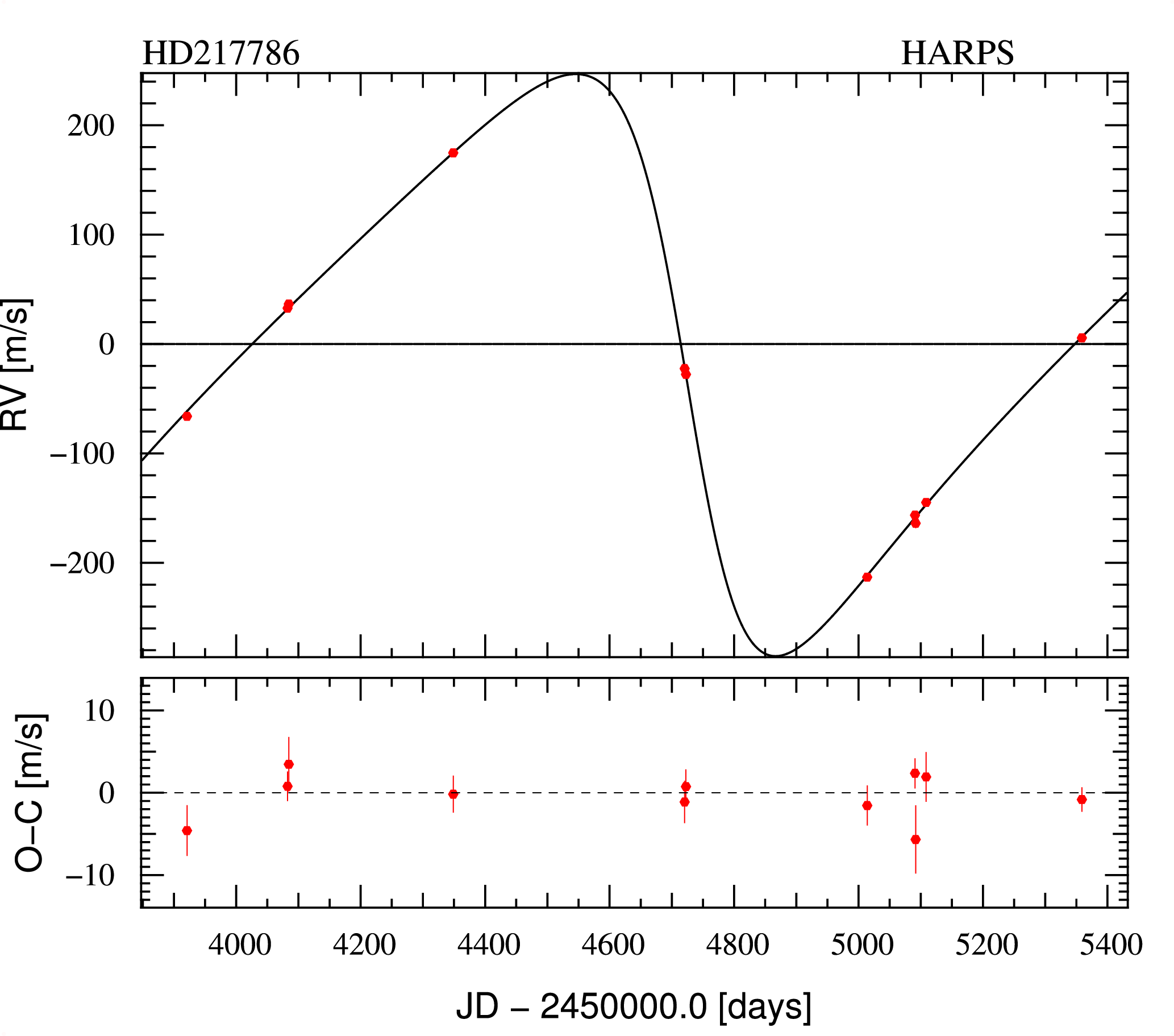,width=8cm}
\caption{Radial-velocity measurements of \i obtained with HARPS against time. }
\label{timei}
\end{figure}

Fifteen spectra of \i were acquired from July 2006 to October 2009. They show very large velocity variations, with a peak-to-peak amplitude of 390 m/s when individual errors are 2.1 m/s on average.  The bisector span shows small variations that are not correlated with the radial velocity, the FWHM and the calcium activity index are not correlated, and the star is not active with $\log R'_{\mathrm{HK}}$ of -4.86. The velocity variation is thus not caused by activity.
A Keplerian solution with a 1320 day period, 0.4 eccentricity and 260 m/s semi-amplitude is the best-fit solution to the data set (Figure \ref{timei}). The residuals to the fit have an $rms$ of 2.5 m/s. In order to further increase the significance of the fit, additional observations around periastron and apastron would be necessary. The minimum mass of the orbiting planet is 13 \mjup, within the overlapping region between giant planets and brown dwarfs. The complete solution is given in Table \ref{sol1}.

%\subsubsection{\k}
%Finally, we obtained 30 measurements of star \k, from July 2006 to June 2010. They show a large-amplitude variation, of 280 m/s peak-to-peak, much greater than individual errors of 1.9 m/s. When fitting a single planet, one ends up with significant scatter of the residuals to the fit (8.9 m/s), which suggests another component, at longer period (Figure \ref{timek}). Having little constraint to a companion of period larger than 1440 days, the time span of our observations, we fit a linear drift in addition to the Keplerian model. Then, a better solution is found, with a scatter of only 3.4 m/s in the residuals. The linear drift is estimated to be 12 m/s/year.  It is consistent with a known stellar companion of \k (also named GJ 676 A); GJ 676 B is an M3-type companion about 3 magnitude fainter and distant by about 800 AU. The inner companion is then found to be a planet with 4.6 \mjup\, minimum mass, slightly eccentric orbit and 1061 day period. The complete solution is given in Table \ref{sol1}.

\subsection{Incomplete orbits}

\subsubsection{\d}
\begin{figure}[h]
\epsfig{file=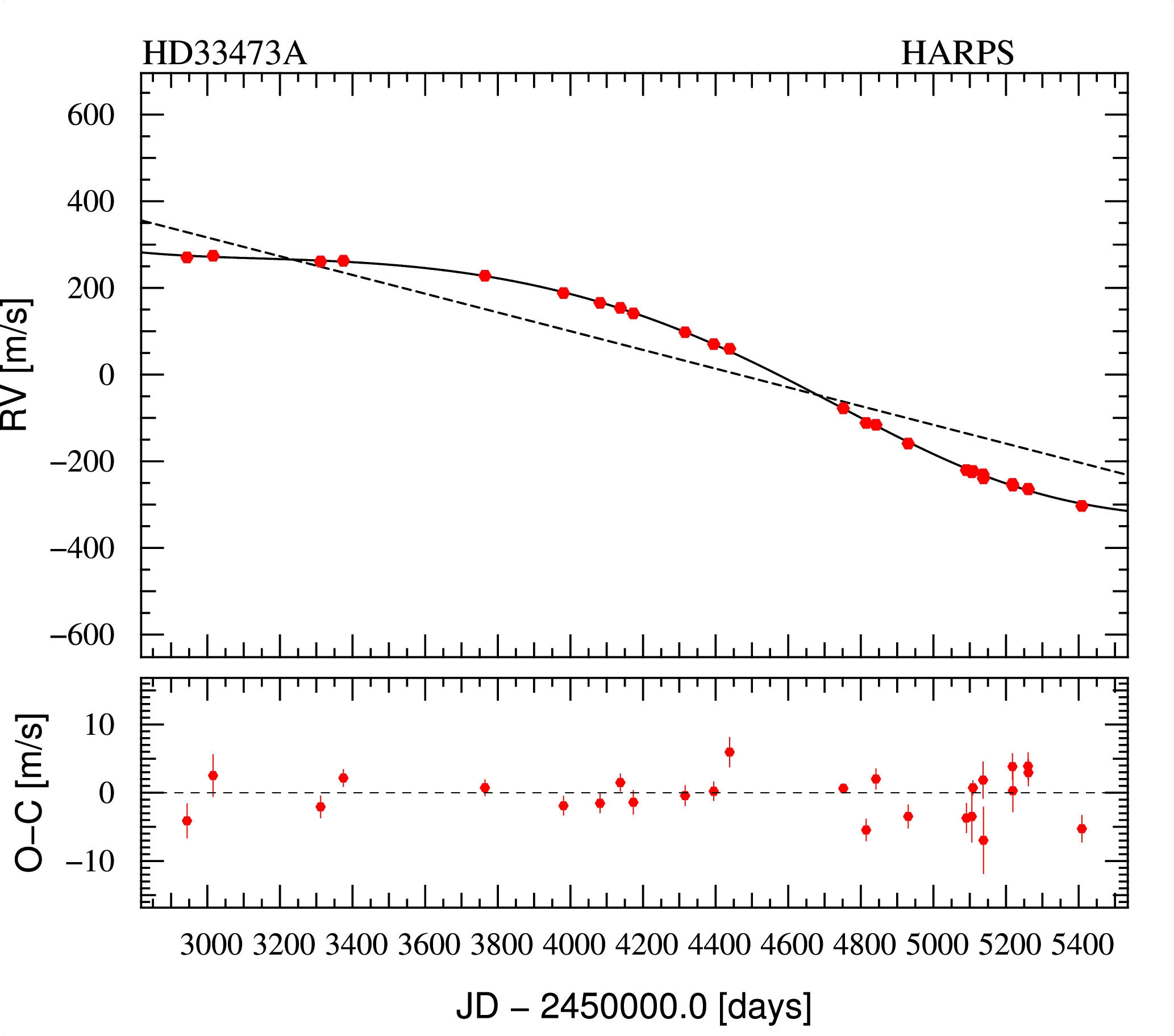,width=8cm}
\epsfig{file=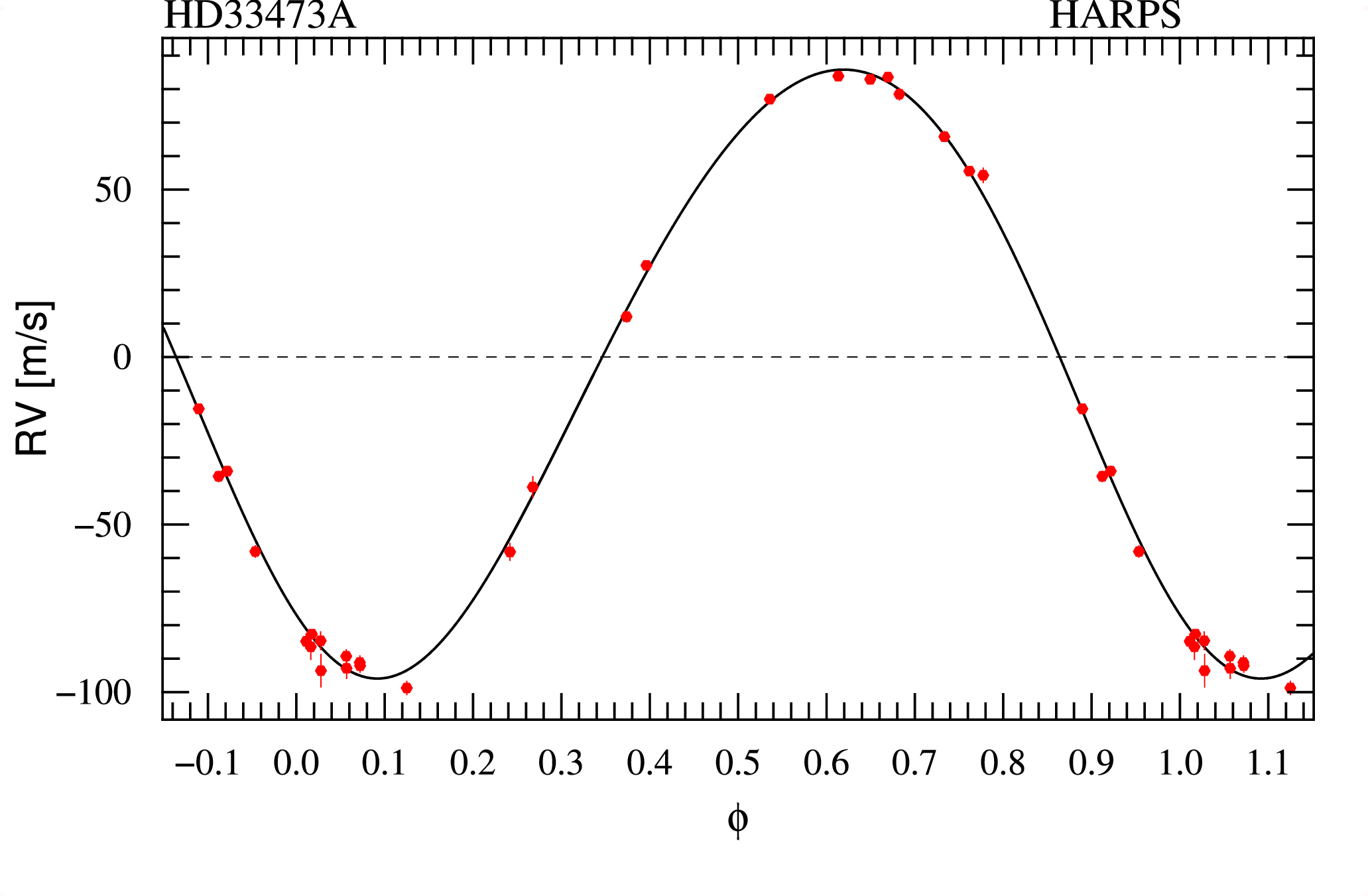,width=8cm}
\caption{Radial-velocity measurements of \d obtained with HARPS against time (left) and phase (right). }
\label{timed}
\end{figure}

Twenty-six HARPS spectra of \d were collected from October 2003 to March 2010 with an average error of 1.9 m/s. Large amplitude variations of the velocity have a standard deviation of 200 m/s. A velocity variation caused by stellar activity is rejected by our data, because the bisector span is constant. 
 The 2300-day time span covered by our observations is not long enough to distinguish between two solutions: an uncomplete, single orbit with an ill-constrained period (16 to 38 years), or a $\sim$2800 day period orbit with a longer-term drift. Both solutions give a similar $\chi^2$. In another year, with continuing observations of this system, we should be able to better constrain the system, because then the shorter period will be entirely observed. With such a difference in the derived period, the mass of the companion is evidently very uncertain, ranging from 6 to 29 \mjup. However, \d has a known visual companion, HD 33473B, with an angular separation of 10 arcsec and magnitude difference 3.4 in the optical. It is thus a K8-like star at 500 UA distance, at least, which corresponds to a period longer than 10$^4$ years. The existence of this distant stellar companion strongly reinforces the justification of a planet model involving a long-term drift. This is the solution presented in this paper (Table \ref{sol2} and Figure \ref{timed}). It gives an $O-C$ residual scatter of about 2.4 m/s. 

%A period even longer than 7500 days is however rejected by our data, so that the mass of the orbiting body remains within the regimes of planets' or brown dwarfs' masses. 
%A preliminary solution involving a planet without a linear drift are presented in Table \ref{sol2} and Figure \ref{timed}; 
%it gives a $O-C$ residual scatter of about 1.9 m/s.  Figure \ref{statd} also shows the correlation plot between period and eccentricity, and the $\chi^2$ map of the secondary minimum mass, obtained by Monte-Carlo Markov-Chain analysis.

\subsubsection{\f}
\begin{figure}[h]
\epsfig{file=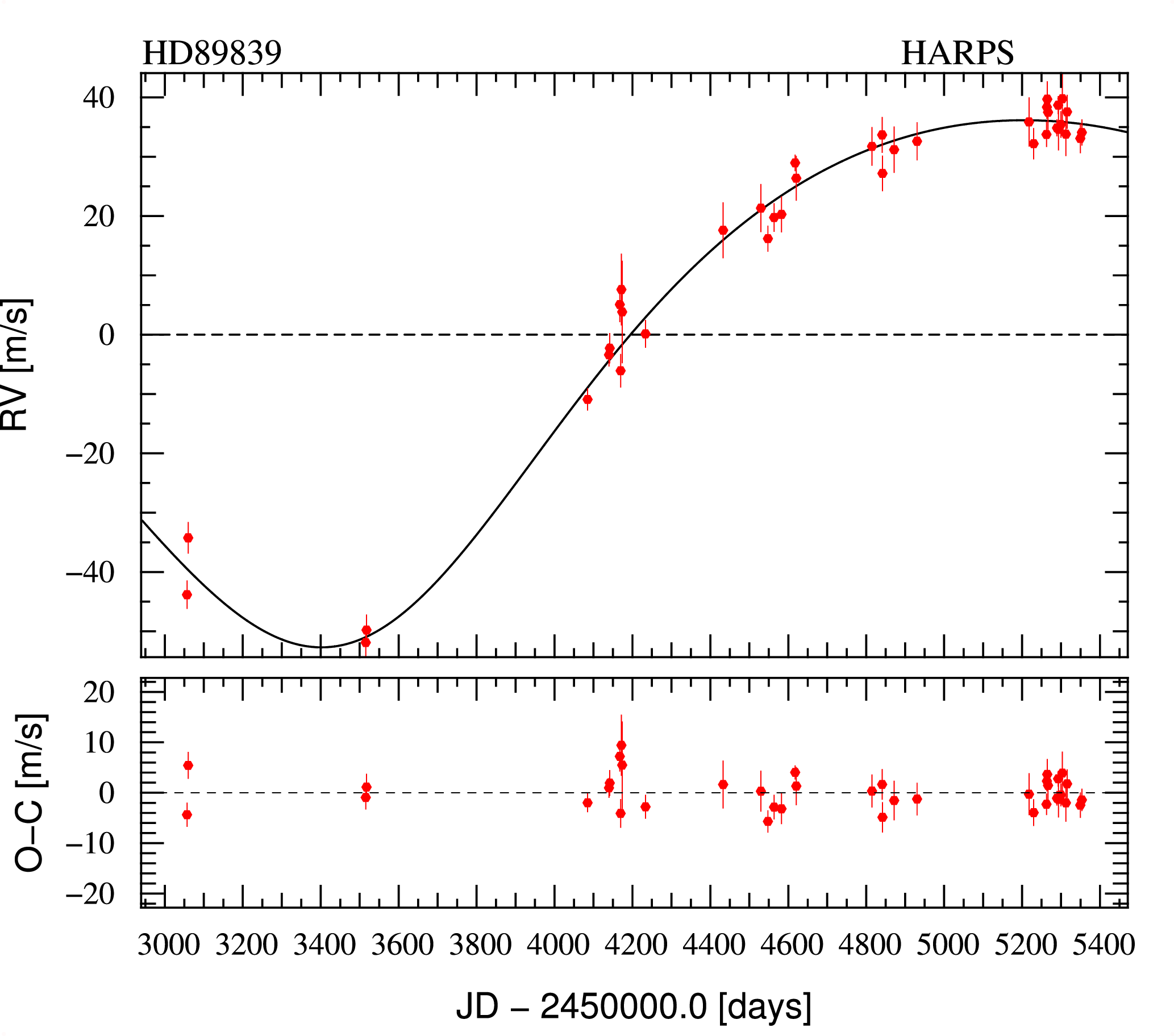,width=8cm}
\caption{Radial-velocity measurements of \f obtained with HARPS against time. }
\label{timef}
\end{figure}

\begin{figure}[h]
\epsfig{file=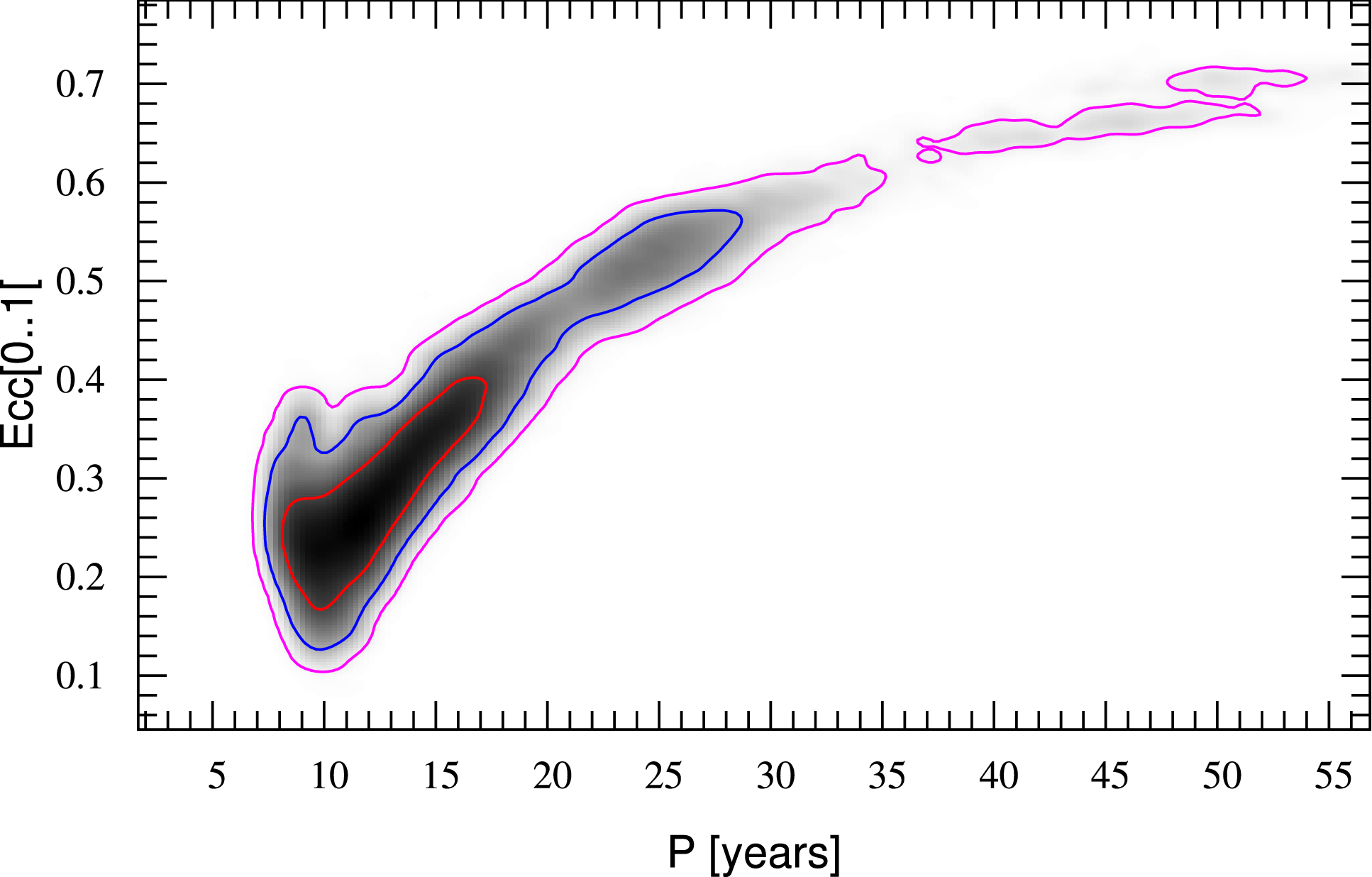,width=8cm}
\epsfig{file=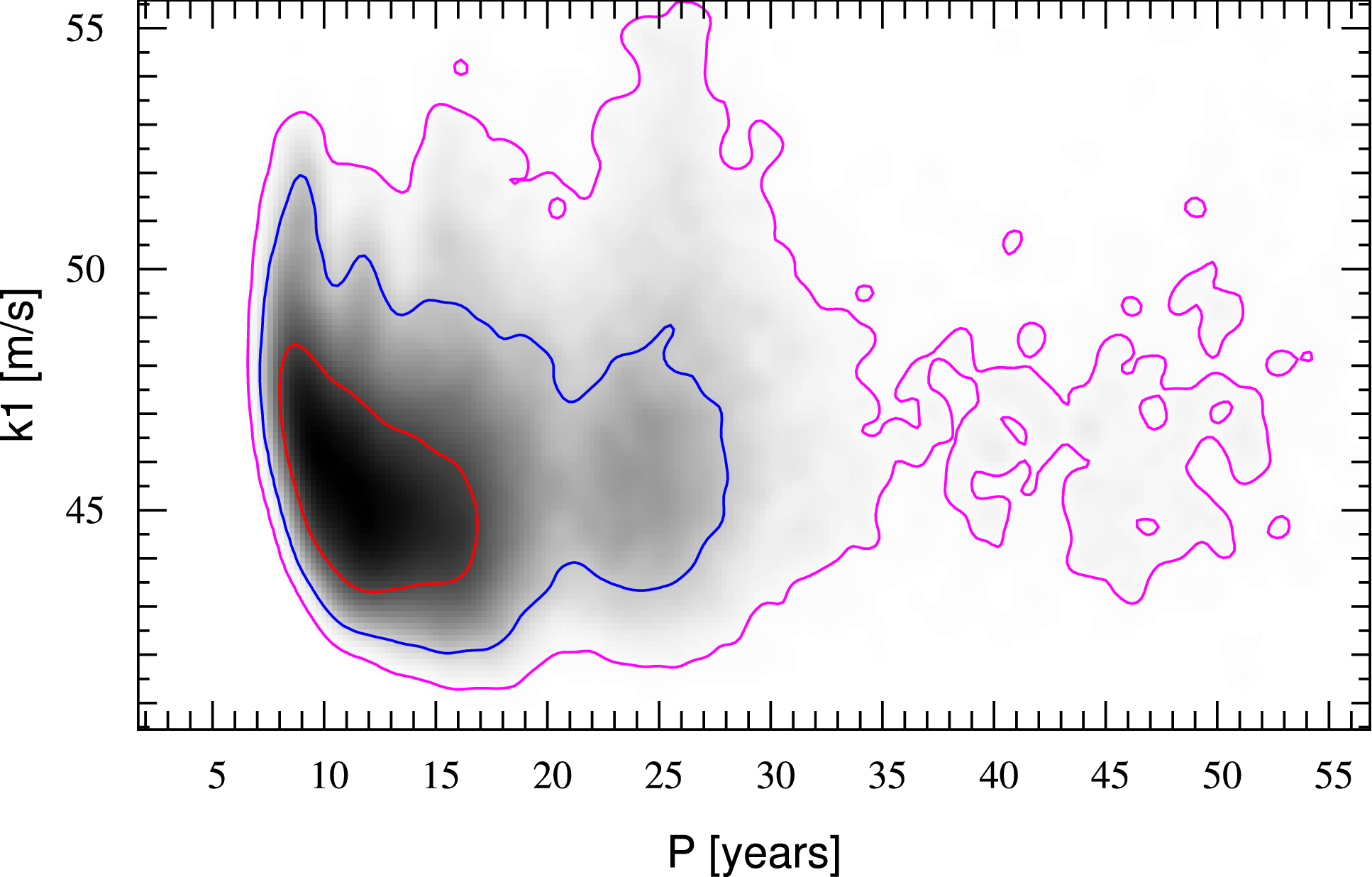,width=8cm}
\caption{Statistical tests on the RV time series of \f: (left) correlation plot between period and eccentricity; (right) correlation plot between velocity semi-amplitude and period.  }
\label{statf}
\end{figure}

From February 2004 to June 2010, 39 measurements of the low-activity star \f were recorded. They show long-term radial-velocity variations with a peak-to-peak amplitude of 90 m/s (Figure \ref{timef}). No correlation is found between the bisector and the velocity, excluding the stellar activity as the origin of the observed signal. Although the orbit is not regularly sampled by the observations, a period of 9.5 to 27 years days emerges, with some excentricity.  As we are entering the descending section of the radial velocity curve, this solution will be confirmed and refined within the next four years. The current solution gives a semi-amplitude variation of 45 m/s, corresponding to a 3.9 \mjup\, planet. The standard deviation of the residuals to the fit is 3.9 m/s, similar to the average error bar of the measurements.  The MCMC output distributions show the correlated behaviour of eccentricity, semi-amplitude, and orbital period in Figure \ref{statf}. 
% By assuming a circular orbit, we increase a little the scatter of the residuals, but we get a higher constraint on the orbital period. This is thus the reference solution, as given in Table \ref{sol2}.

\subsubsection{\h}
\begin{figure}[h]
\epsfig{file=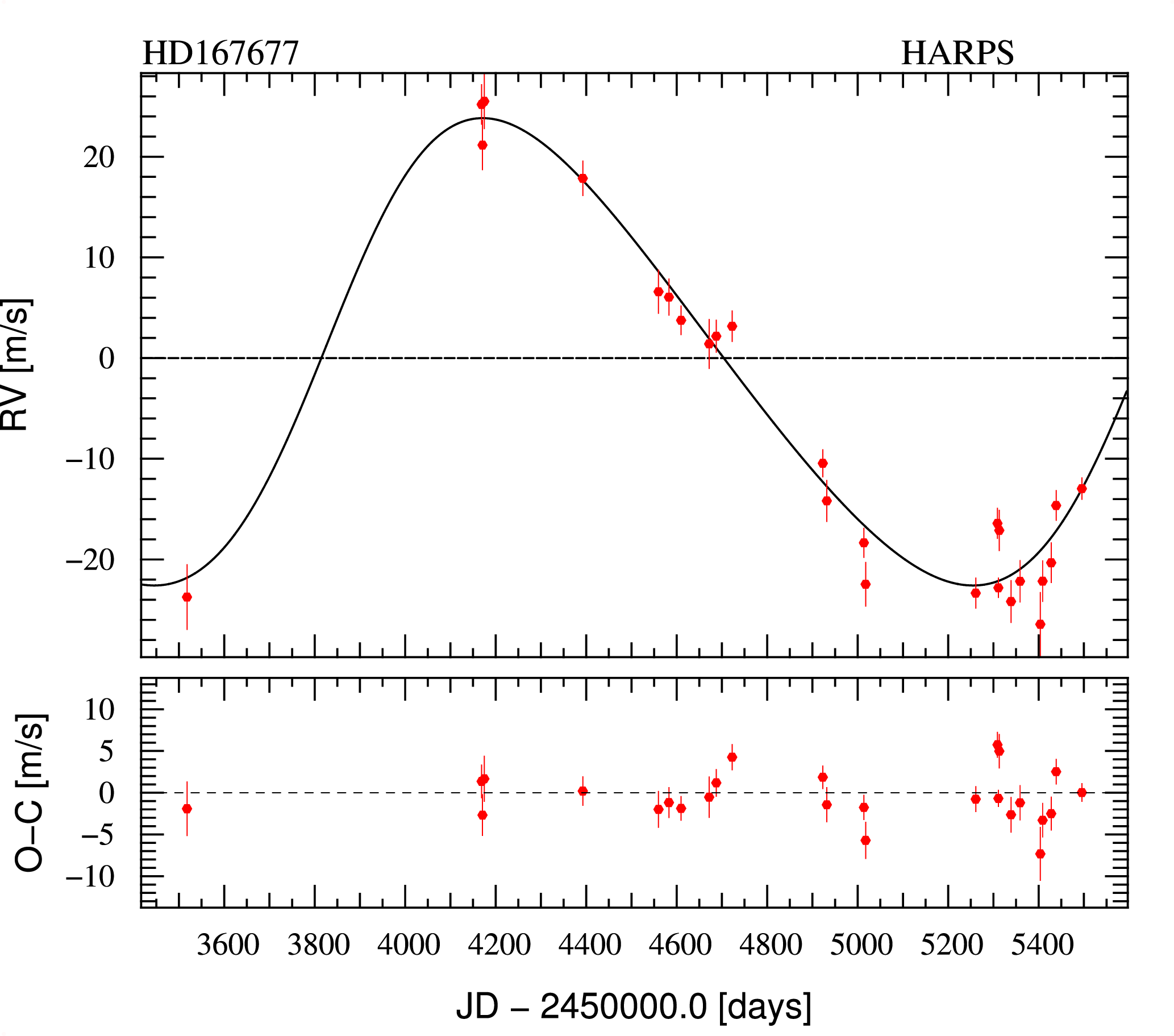,width=8cm}
\caption{Radial-velocity measurements of \h obtained with HARPS against time. }
\label{timeh}
\end{figure}

\begin{figure}[h]
\epsfig{file=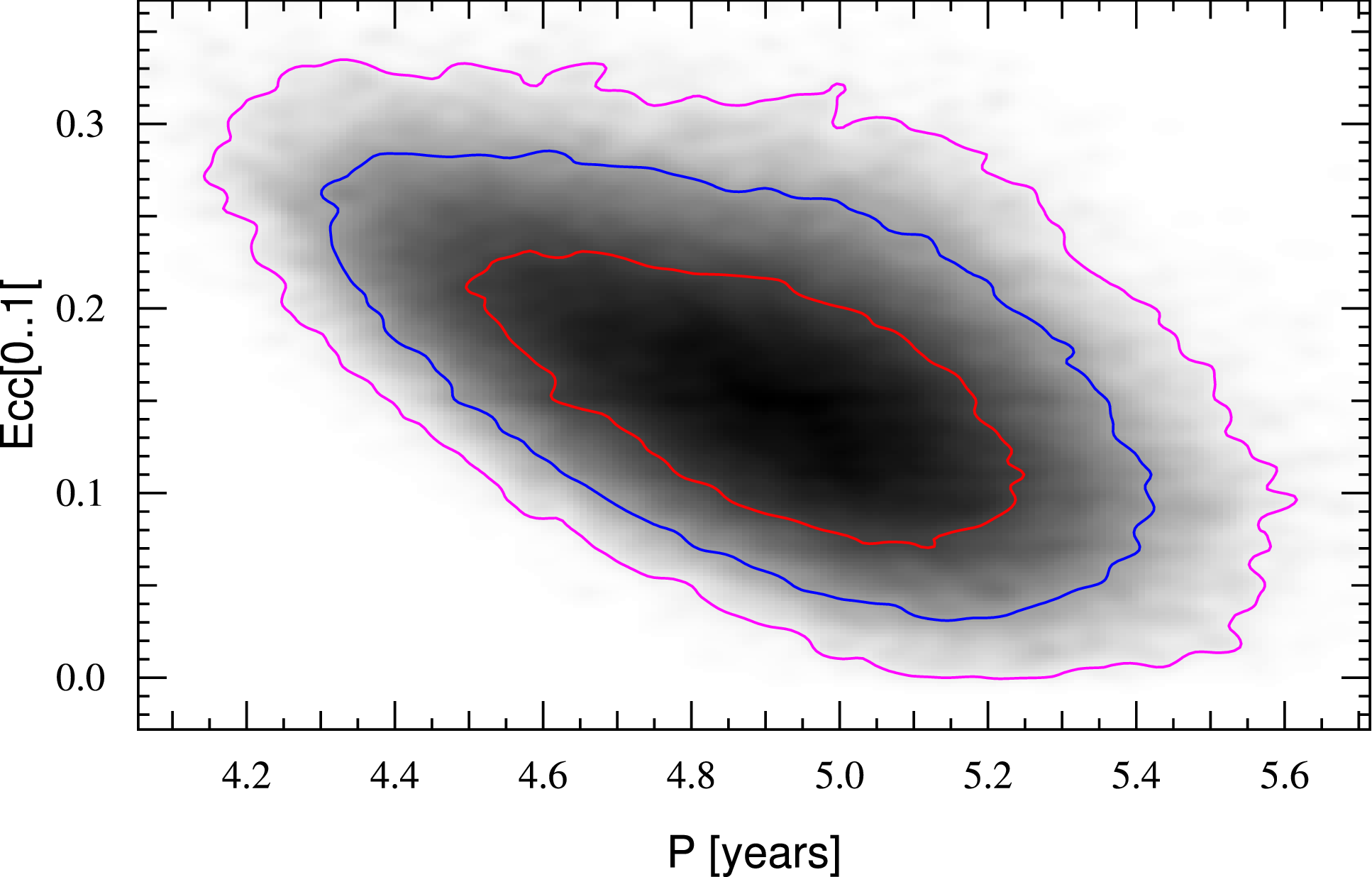,width=8cm}
\epsfig{file=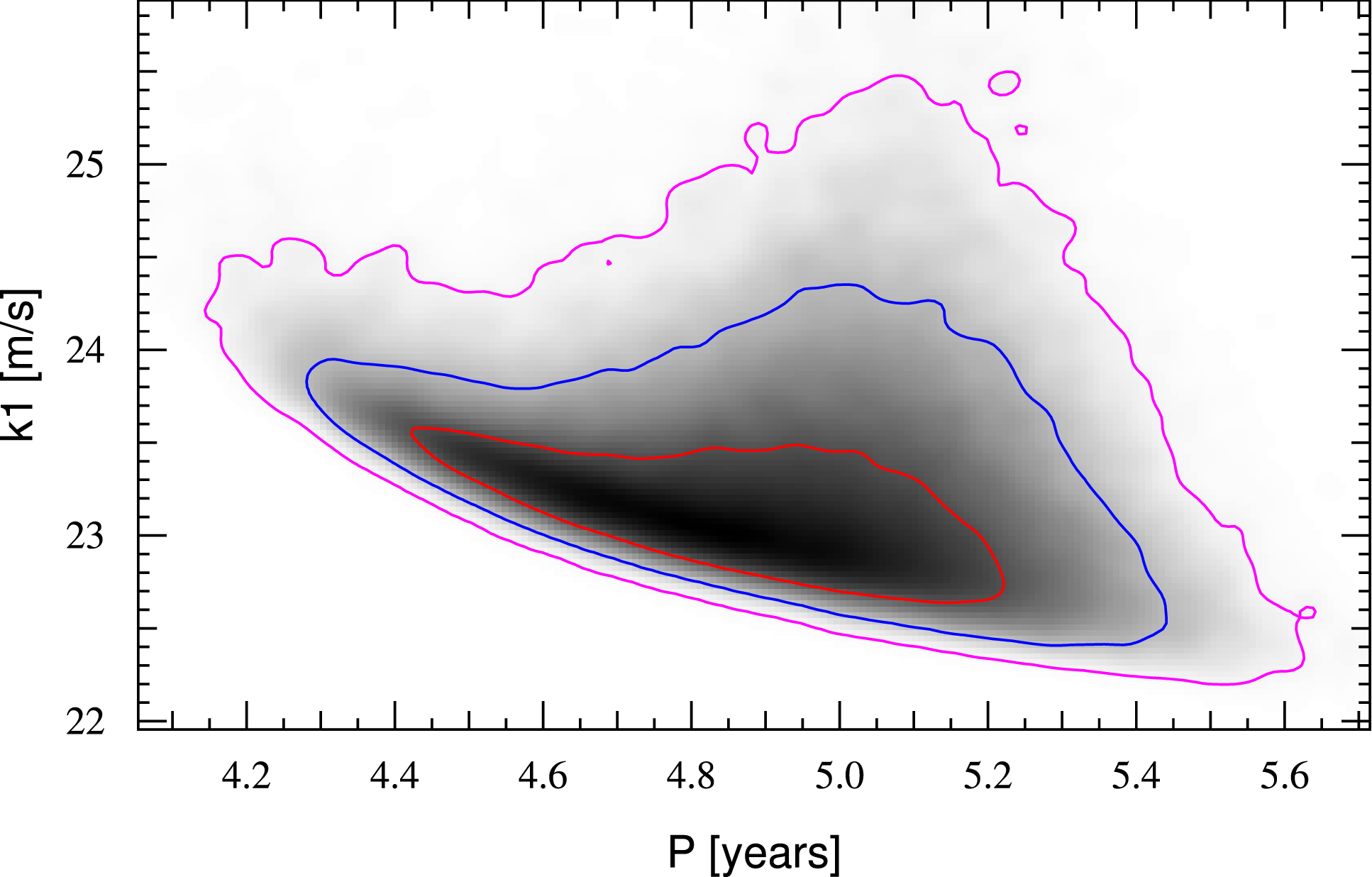,width=8cm}
\caption{Statistical tests on the RV time series of \h: (left) correlation plot between period and eccentricity; (right) correlation plot between velocity semi-amplitude and period.  }
\label{stath}
\end{figure}

Twenty-six HARPS measurements of \h were collected between May 2005 and October 2010, with an average 1$\sigma$ error of 1.6 m/s. They show radial-velocity fluctuations at the level of 15 m/s standard deviation, and with a periodicity of about 1800 days. 
The activity indicator $\log R'_{\mathrm{HK}}$ is typical of a non-active star (-4.99). The bisector span and FWHM of the cross-correlation function do not vary in phase with the calcium index or the velocity. So the interpretation in terms of pure stellar activity is excluded. 
The best-fit Keplerian solution corresponds to a planet with minimum mass 1.36 \mjup\, in a slightly eccentric orbit, as shown in Figure \ref{timeh}. The residuals to the fit have a standard deviation of 2.6 m/s.  The MCMC analysis allowed us to quantify the correlation between the orbital parameters, and Figure \ref{stath} shows the relation between eccentricity, period, and semi-amplitude. Additionnal data would be needed to get more precise parameters  of the orbit, and can be collected in the next years. The complete solution is given in Table \ref{sol2}.

\subsection{Long-period planets or magnetic cycles?}
\subsection{\b}
Nineteen HARPS measurements were obtained of star \b between July 2005 and June 2010, with an average error of 1.9 m/s (Figure \ref{timecy}). A peak-to-peak velocity variation of 28 m/s is observed. 
The bisector slope shows a significant degree of anti-correlation (Fig. \ref{allbis}), and a visible reversal in the core of the calcium lines confirms that the star is active. However, the main radial-velocity variation has a significant peak of periodicity at 1692 days, much too long for a rotation period, and no significant peak in the range 5-50 days.  An attempt has been made to correct for the activity as in \citet{melo} and \citet{boisse09} by linearly fitting the bisector-velocity correlation and subtracting this trend from the velocity measurements. The resulting solution is not significantly different. In addition, Figure \ref{cycle}Ê shows a significant degree of correlation between the full-width-at-half maximum of the cross-correlation function, and the calcium activity index (Mount Wilson $S$ value). The stellar lines therefore appear distorted with a long timescale. This behaviour may indicate a velocity variation owing to a slow evolution of the stellar activity, rather than a long-period planetary companion. Both interpretations are possible however, because the degree of correlation as shown in Figure \ref{cycle} is not high (72\%).

When we fit the data in search for a Keplerian orbit, we find a circular orbit of a planet with a minimum mass of 0.6 \mjup\, and 2.5 AU semi-major axis.  It corresponds to a signal at a period of 1500 to 1900 d, or 4 to 5 years. This is compatible with the typical duration of an activity cycle, although on the short end (see e.g. \citet{baliunas95, fares09, metcalfe10}).

\subsection{\j}
\begin{figure}[h]
\epsfig{file=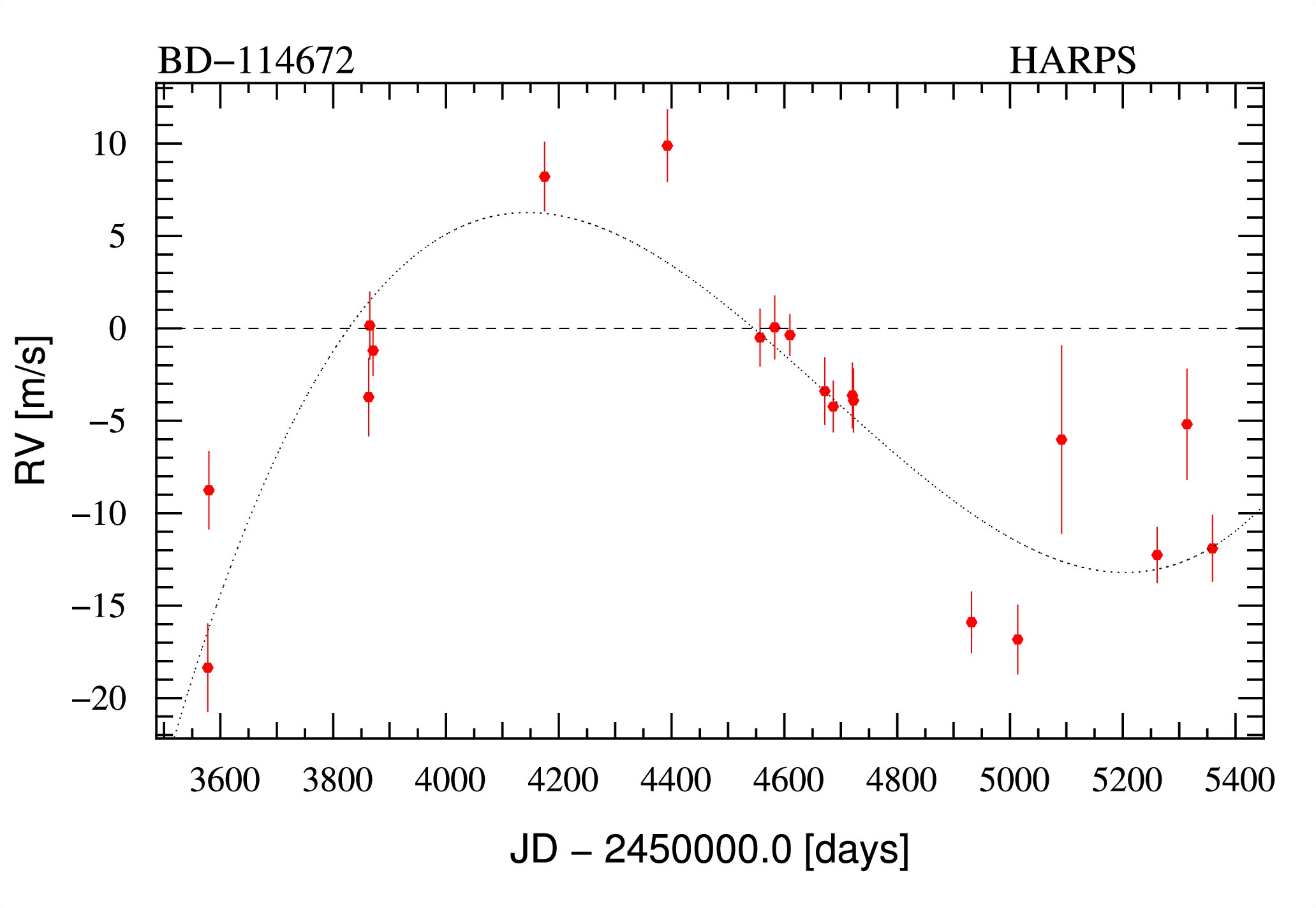,width=8cm}
\epsfig{file=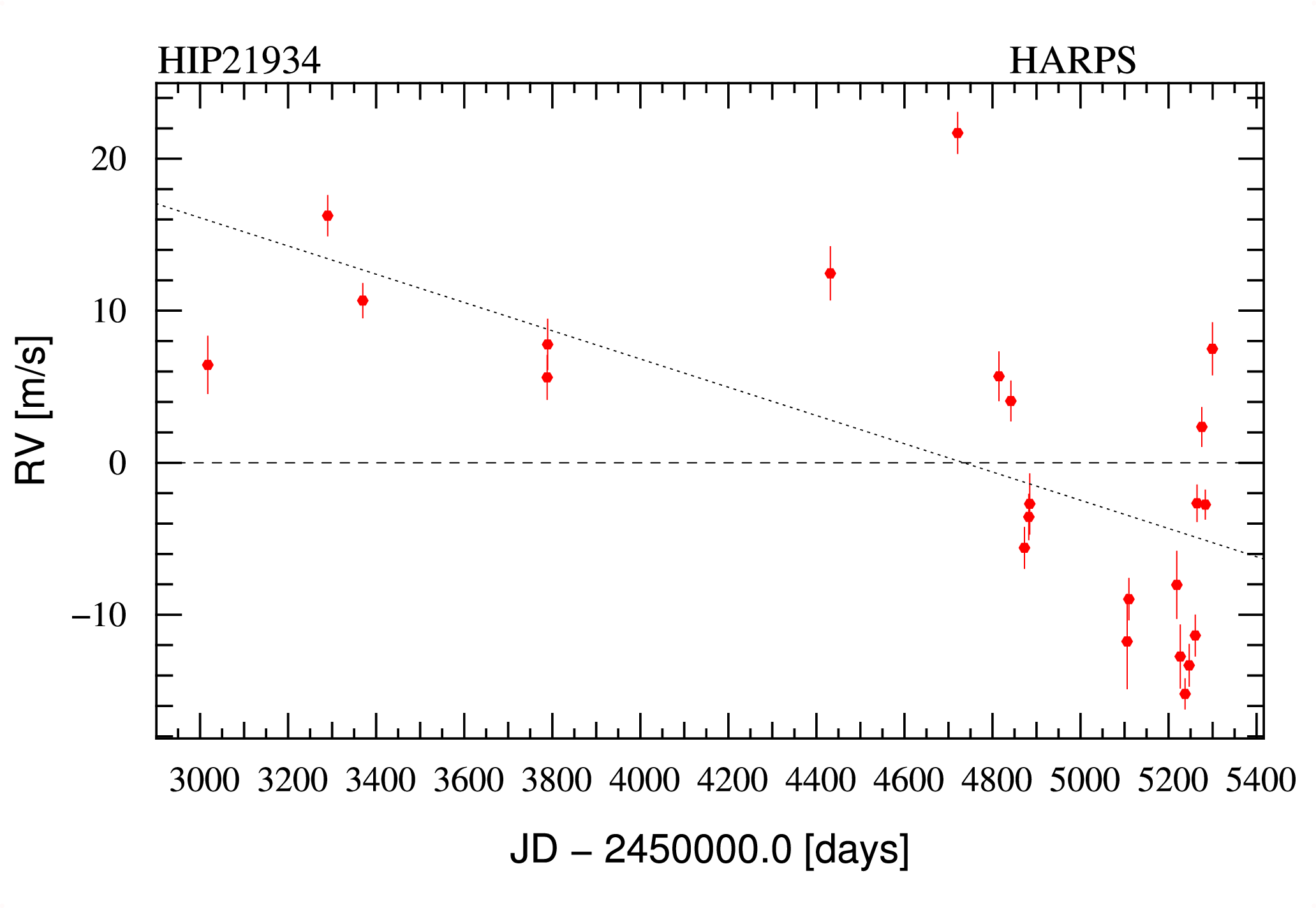,width=8cm}
\caption{Radial-velocity measurements against time of stars \b and \j for which the long-term velocity variation is interpreted as caused by the magnetic activity cycle of the star. }
\label{timecy}
\end{figure}

\begin{figure}
\epsfig{file=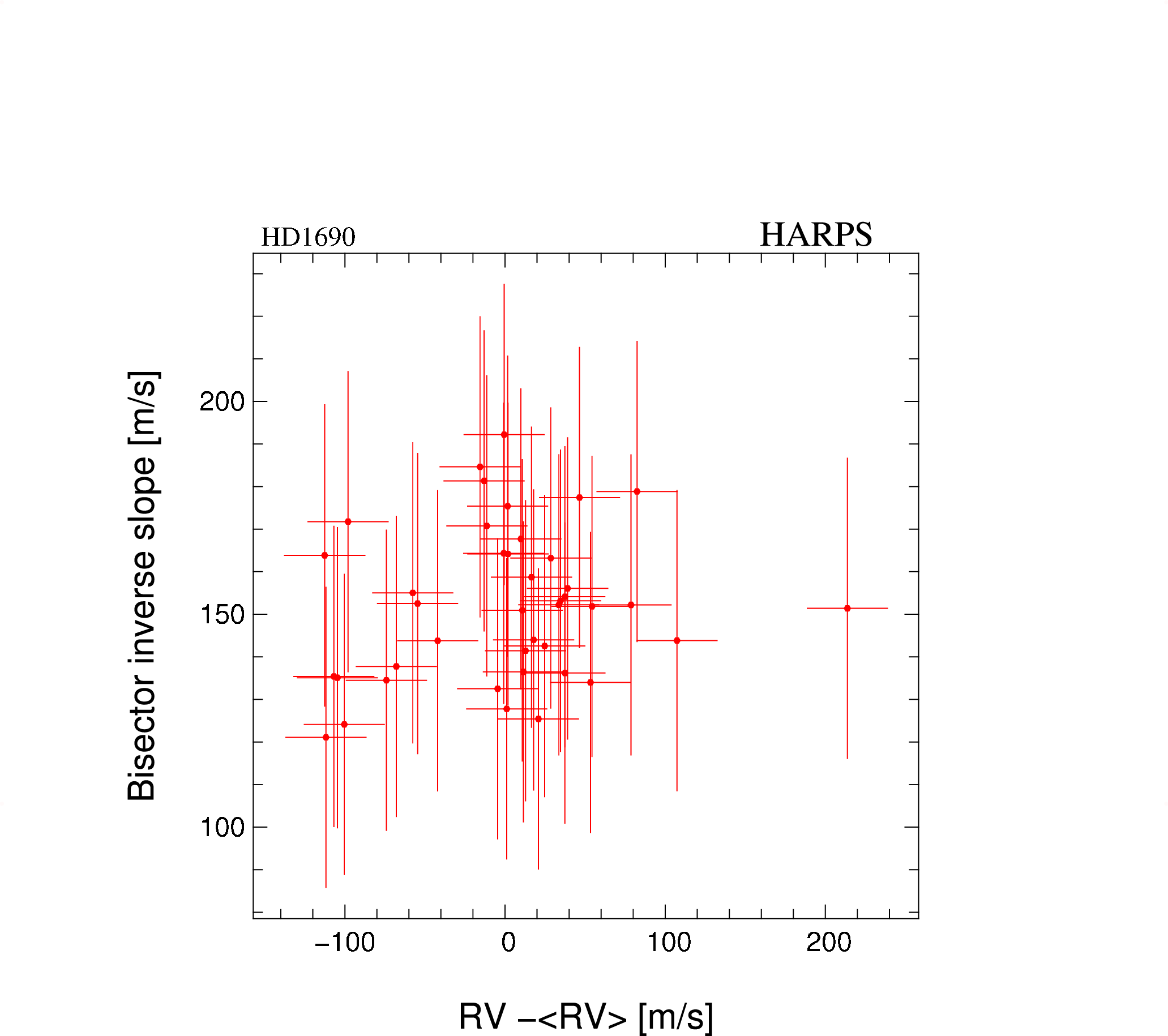,width=5cm}
\epsfig{file=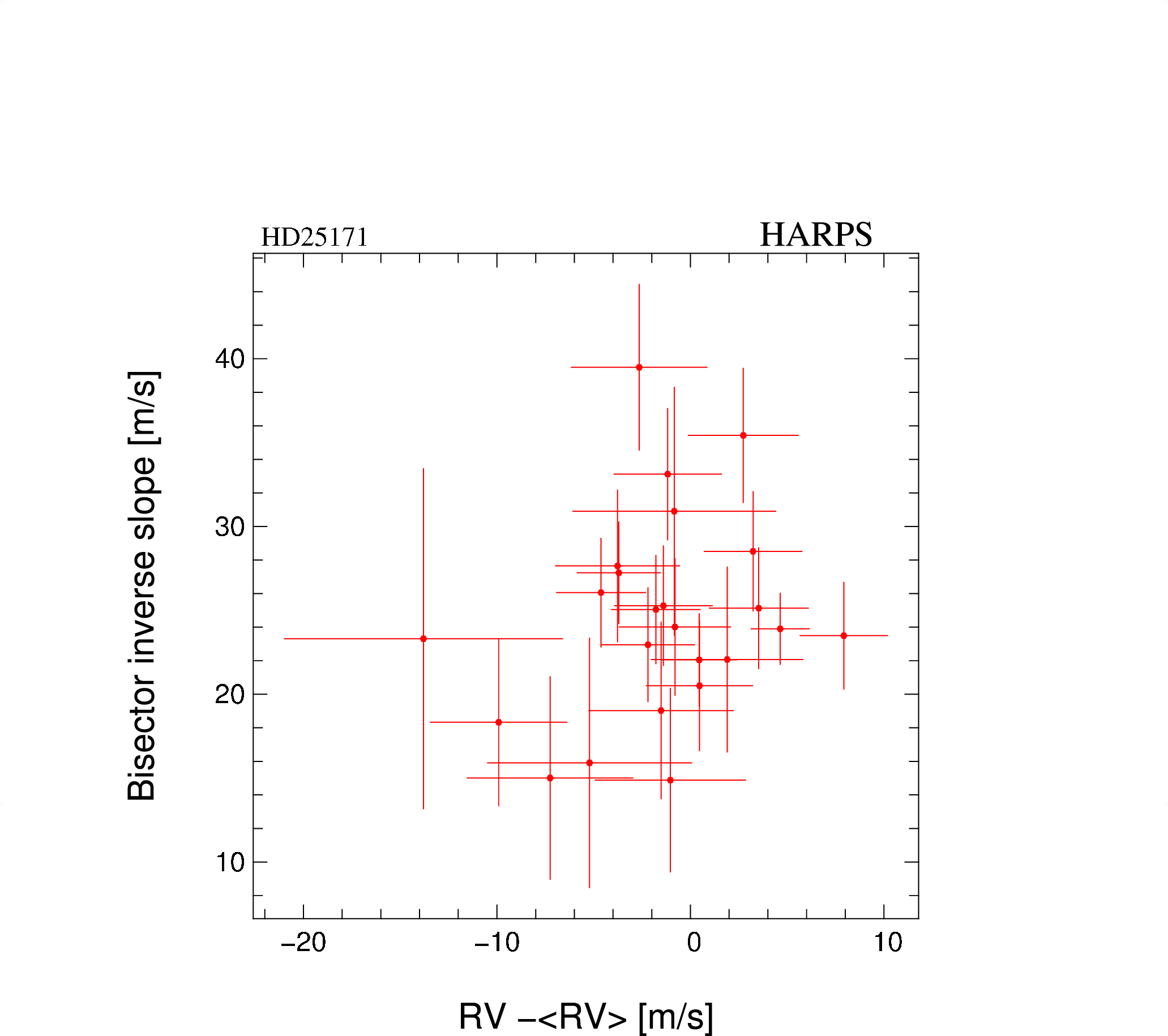,width=5cm}
\epsfig{file=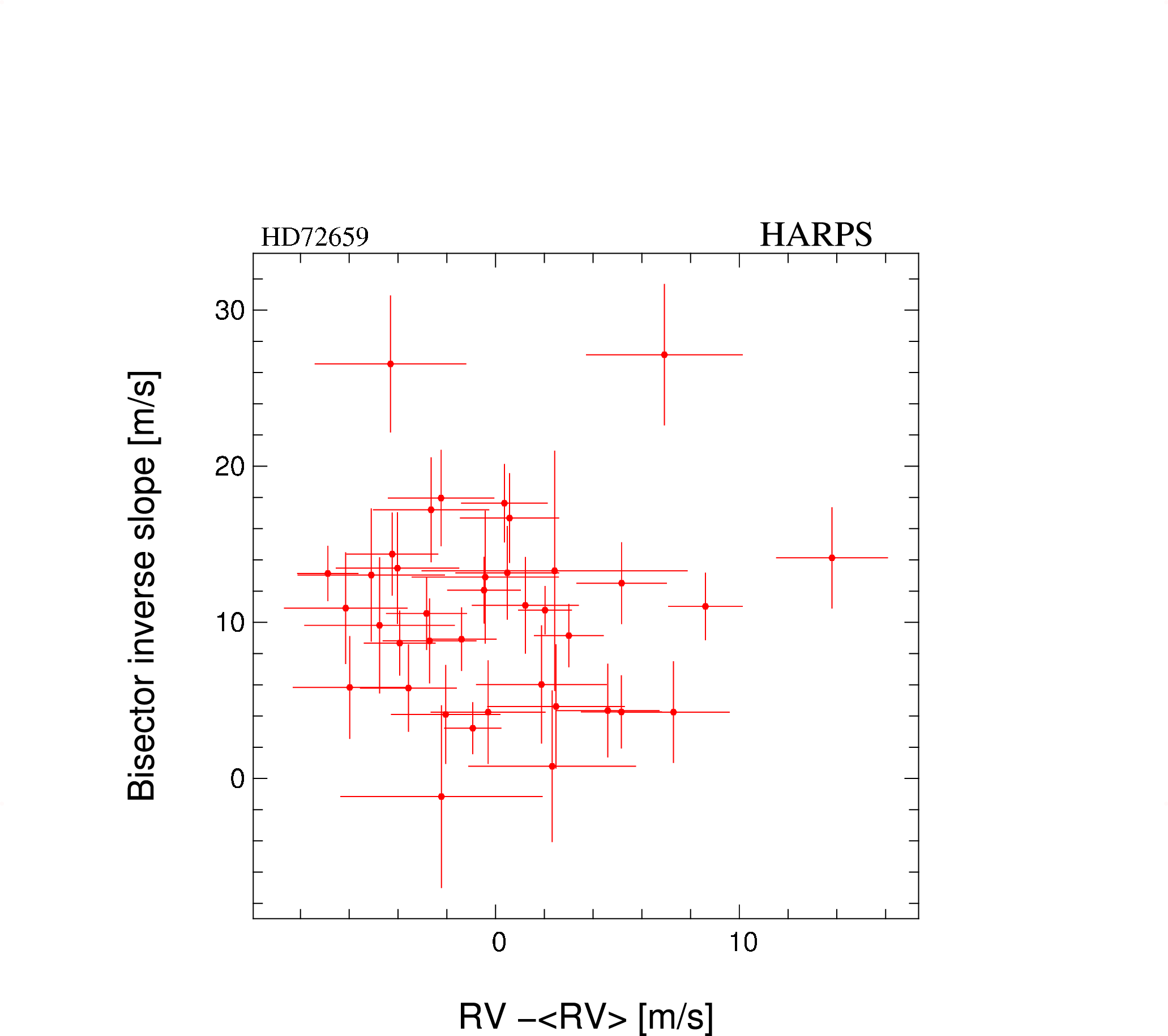,width=5cm}
\epsfig{file=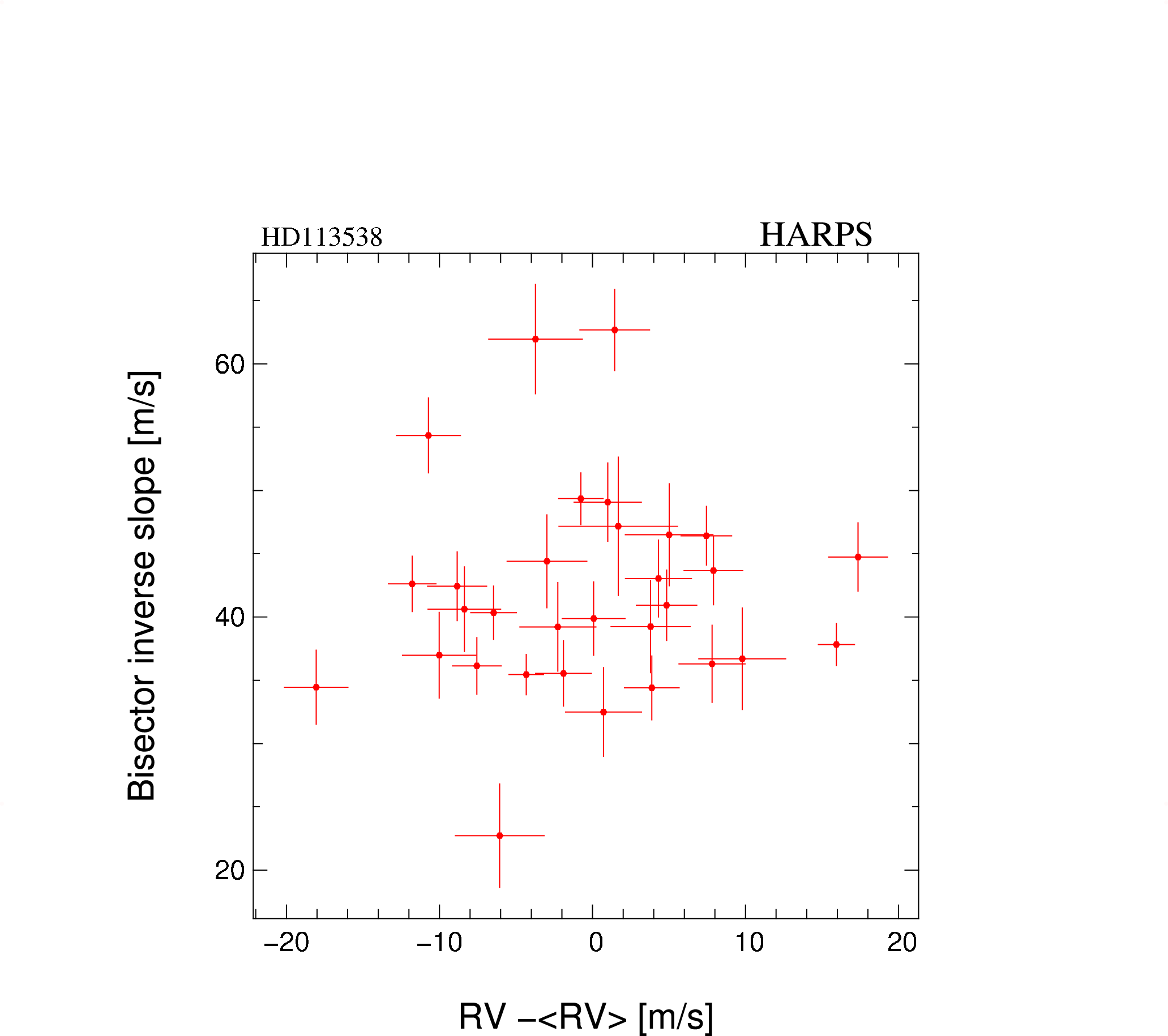,width=5cm}
\epsfig{file=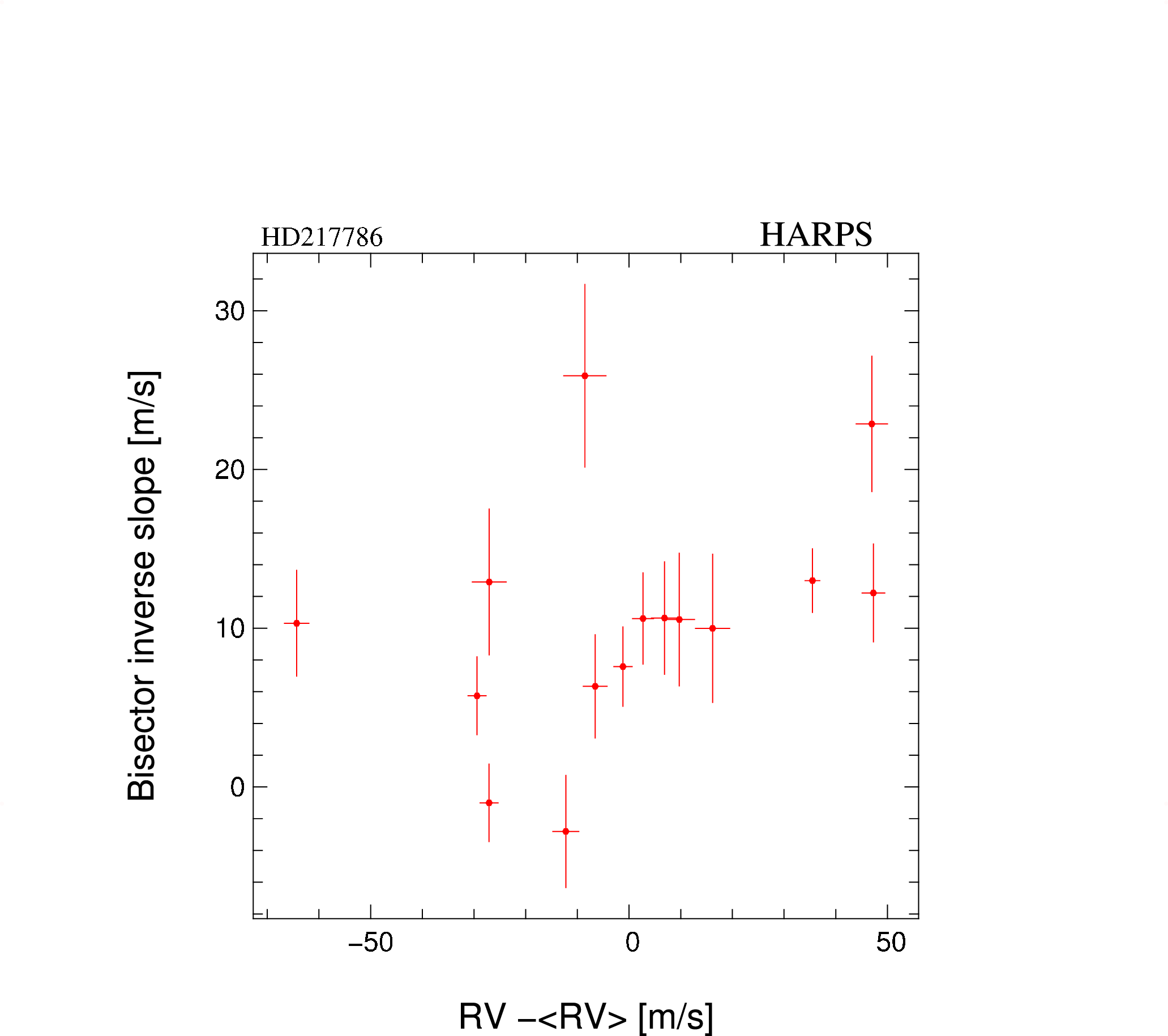,width=5cm}
\epsfig{file=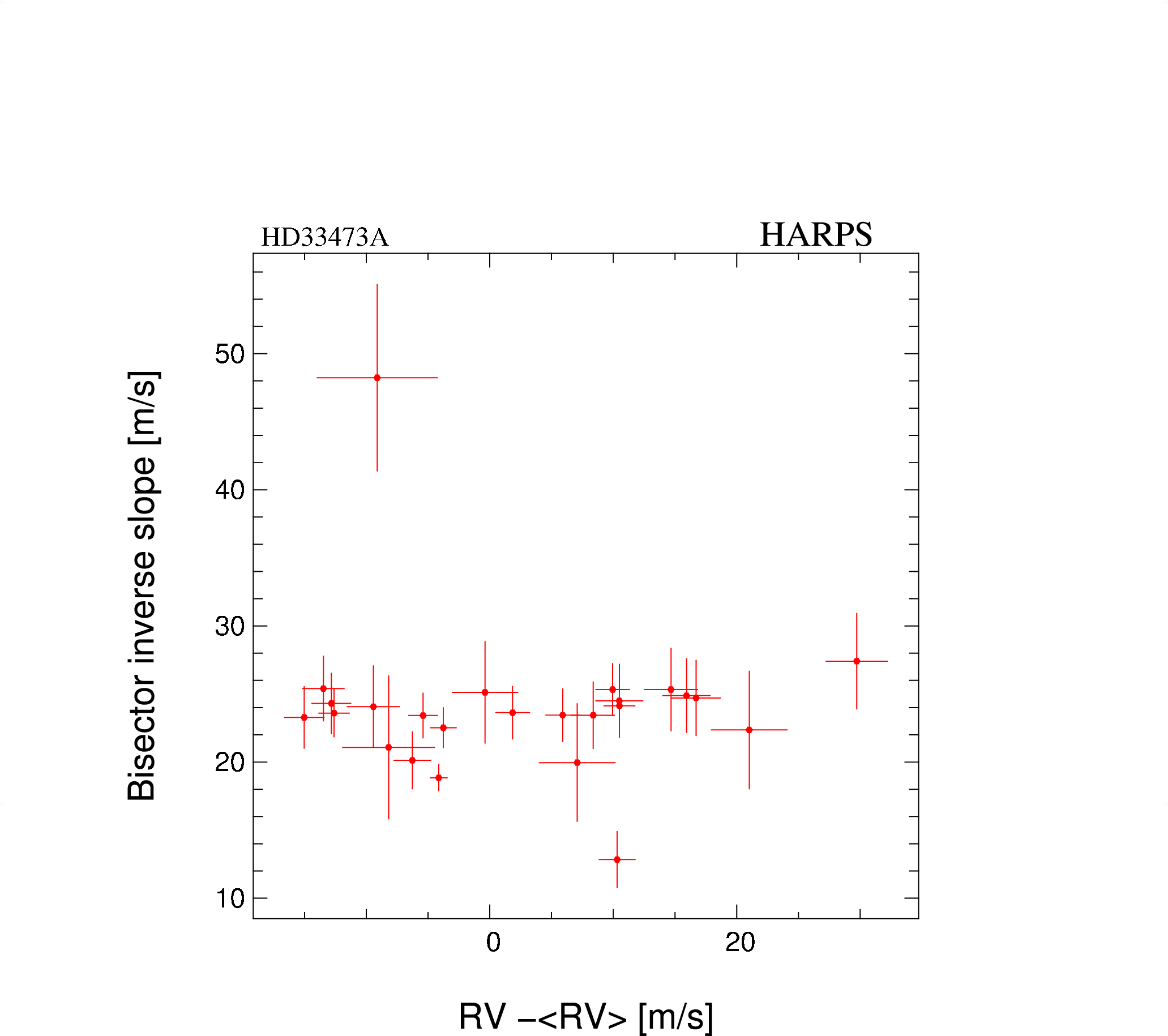,width=5cm}
\epsfig{file=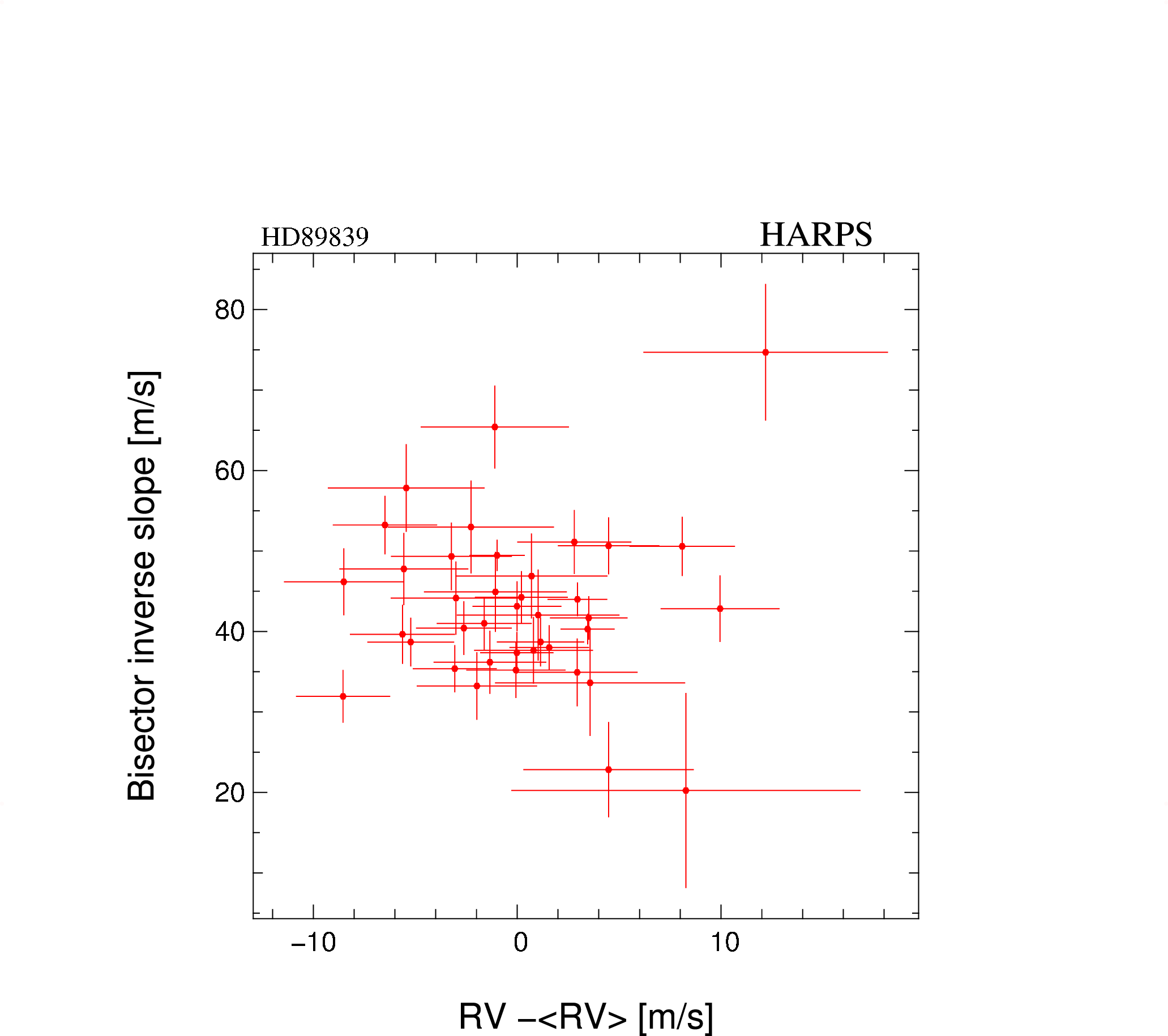,width=5cm}
\epsfig{file=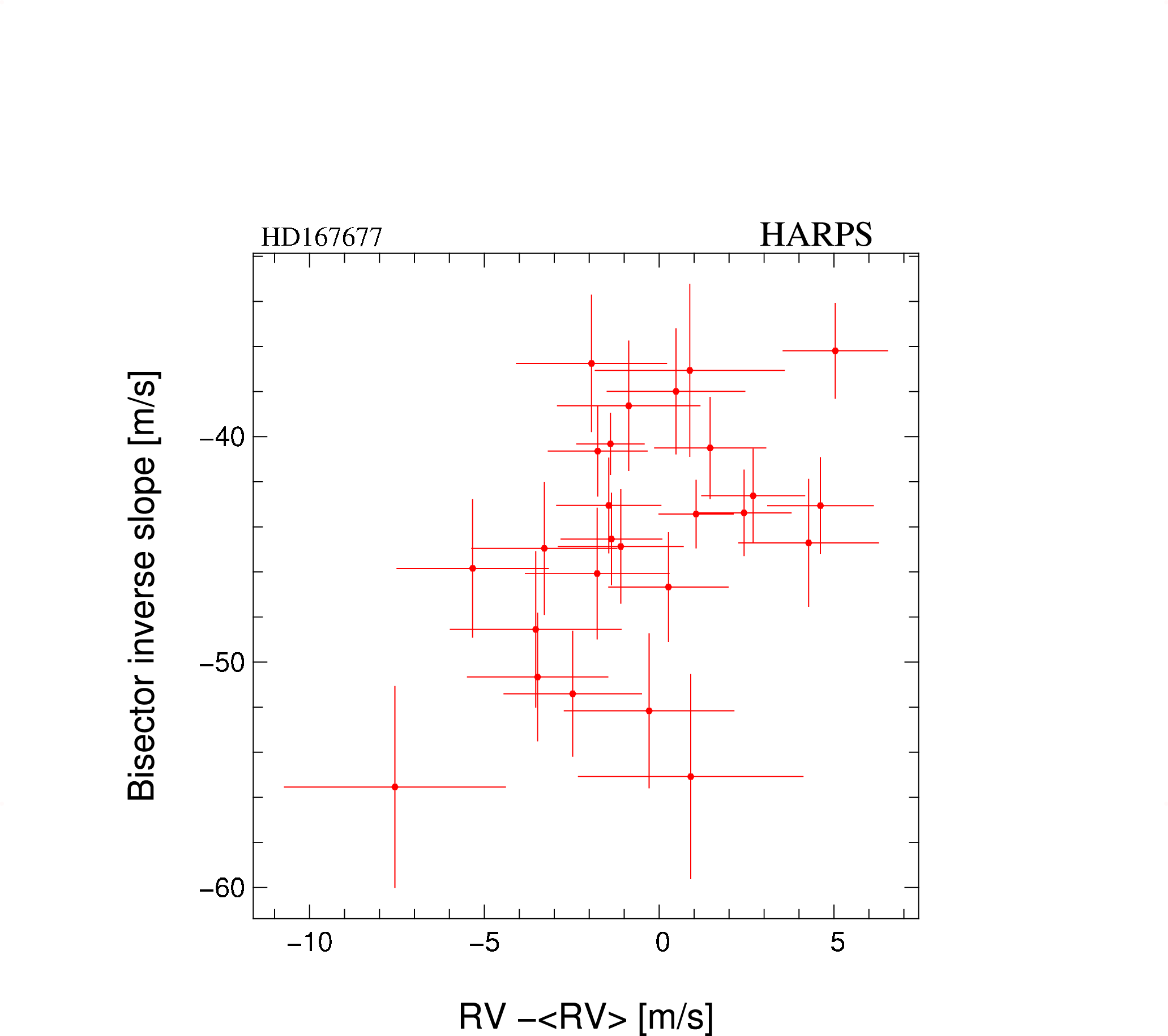,width=5cm}
\epsfig{file=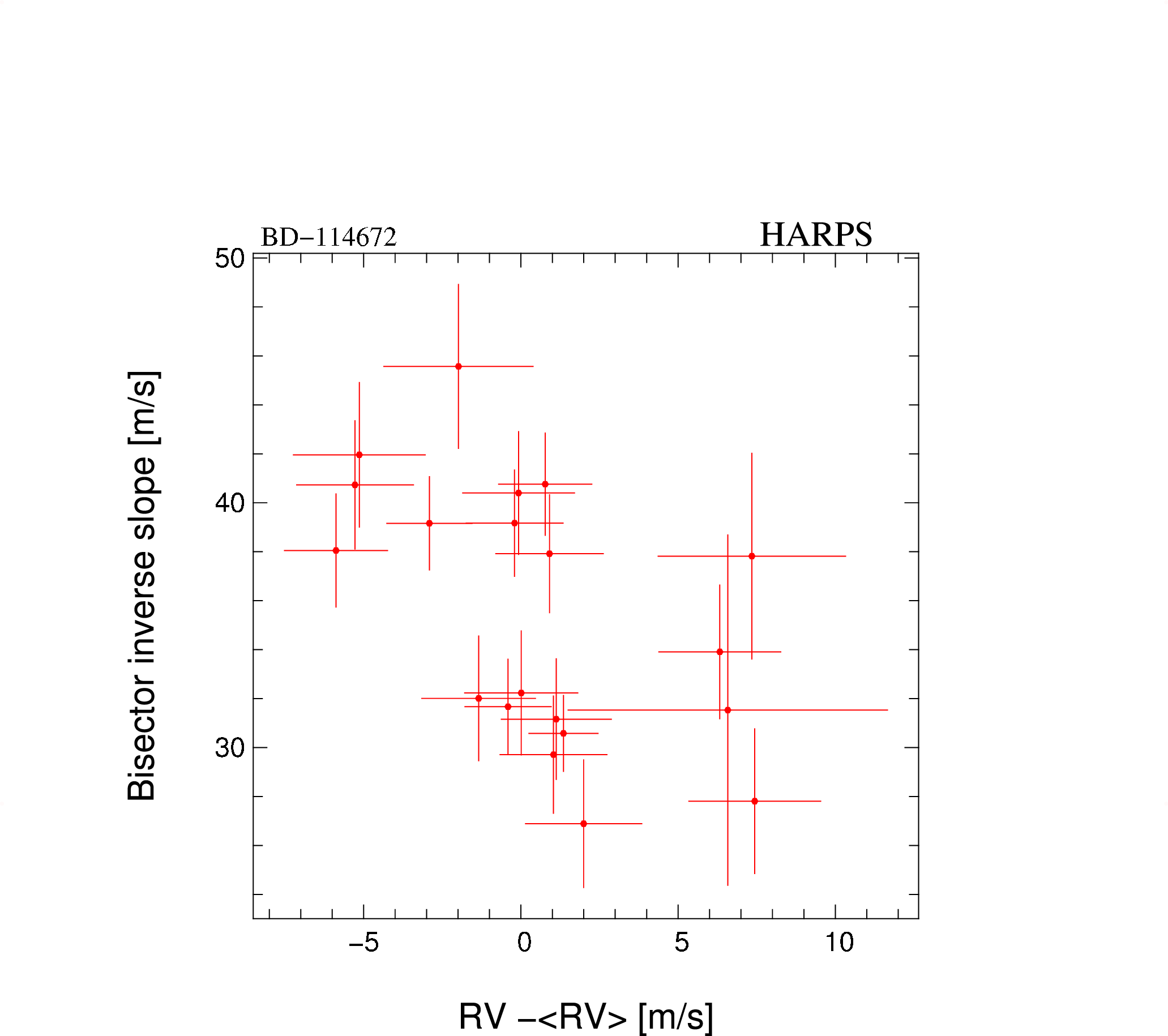,width=5cm}
\epsfig{file=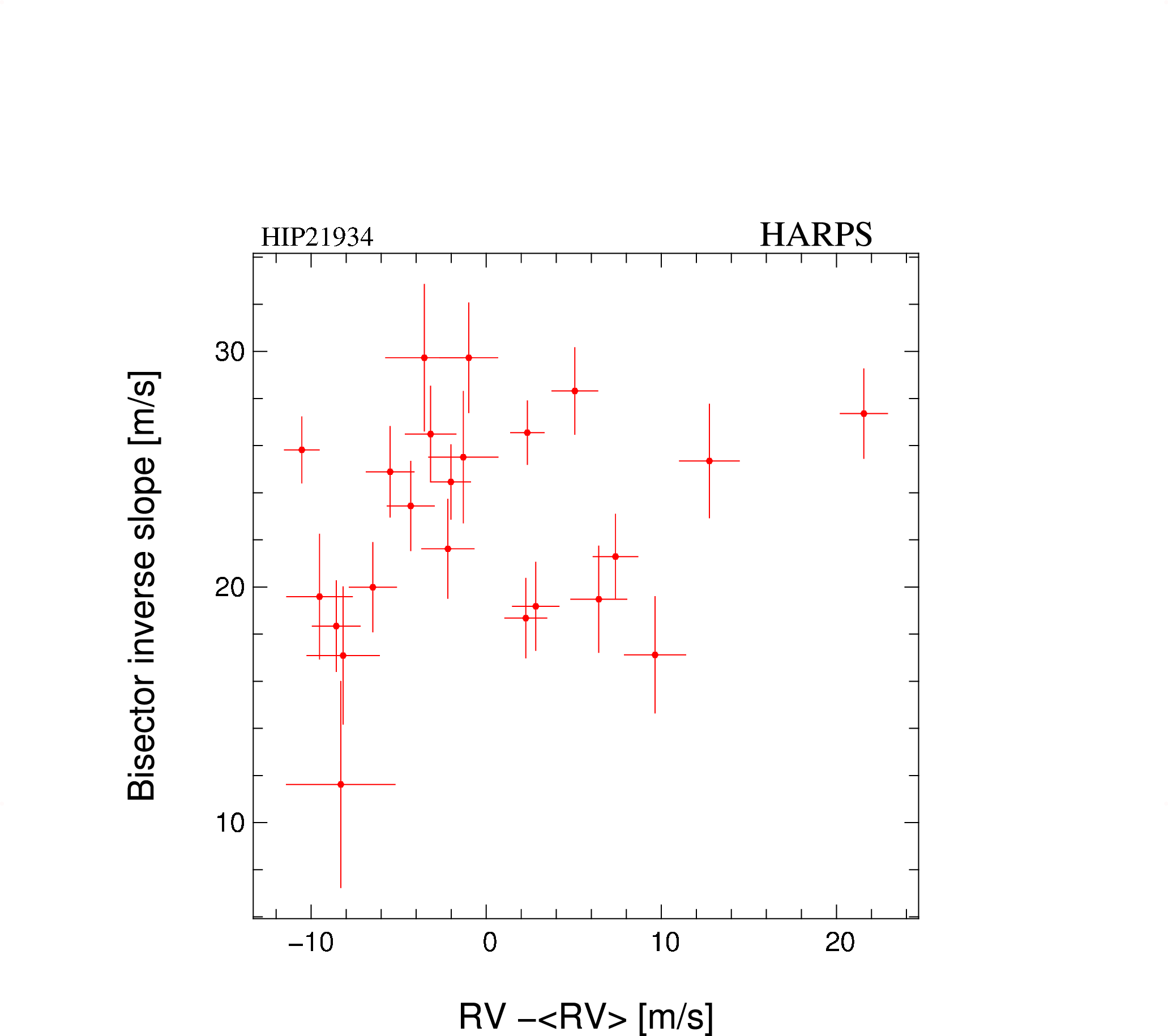,width=5cm}
\caption{Bisector slope variations with respect to radial velocity for all data sets.}
\label{allbis}
\end{figure}

\begin{figure}
\epsfig{file=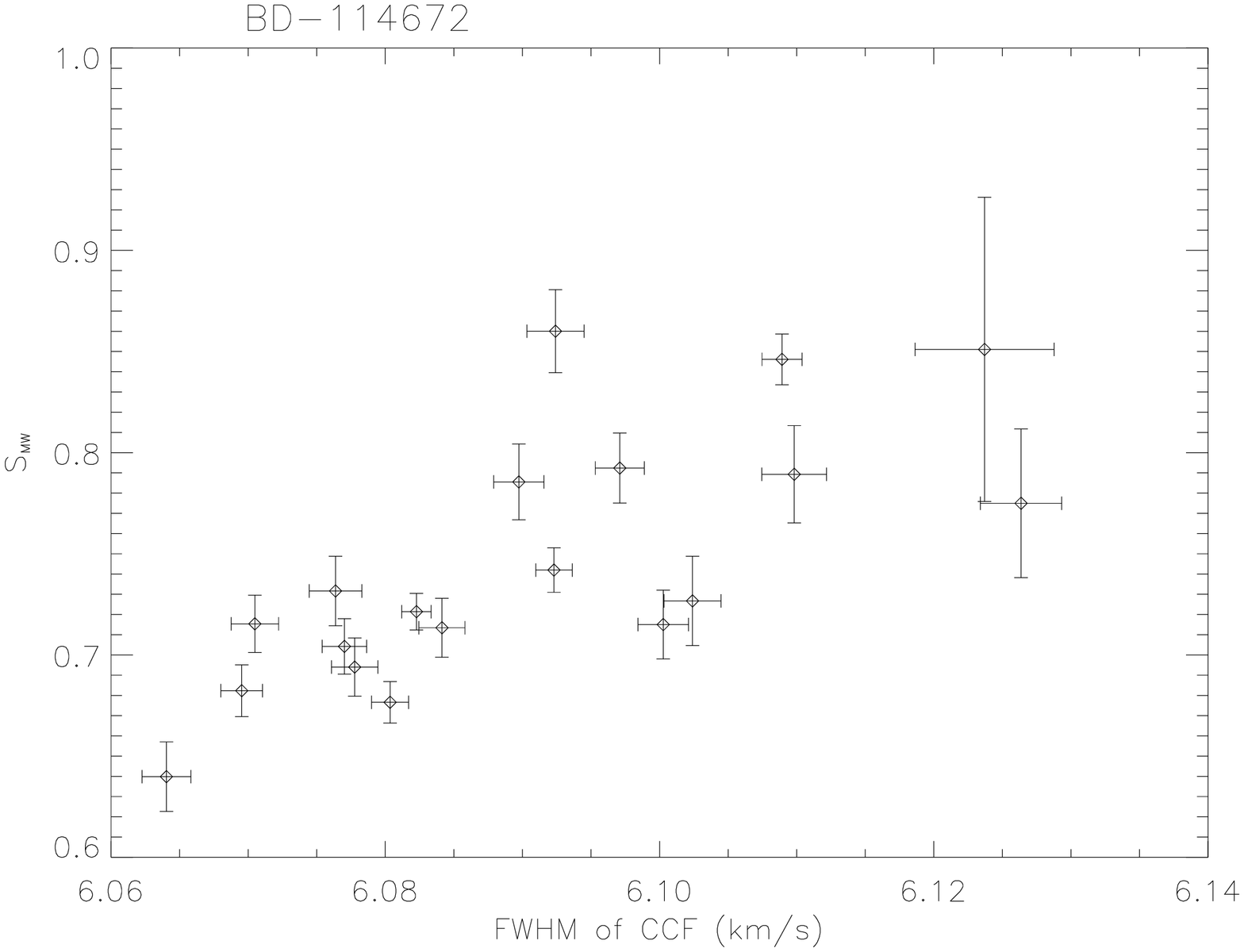,width=6cm,angle=90}
\epsfig{file=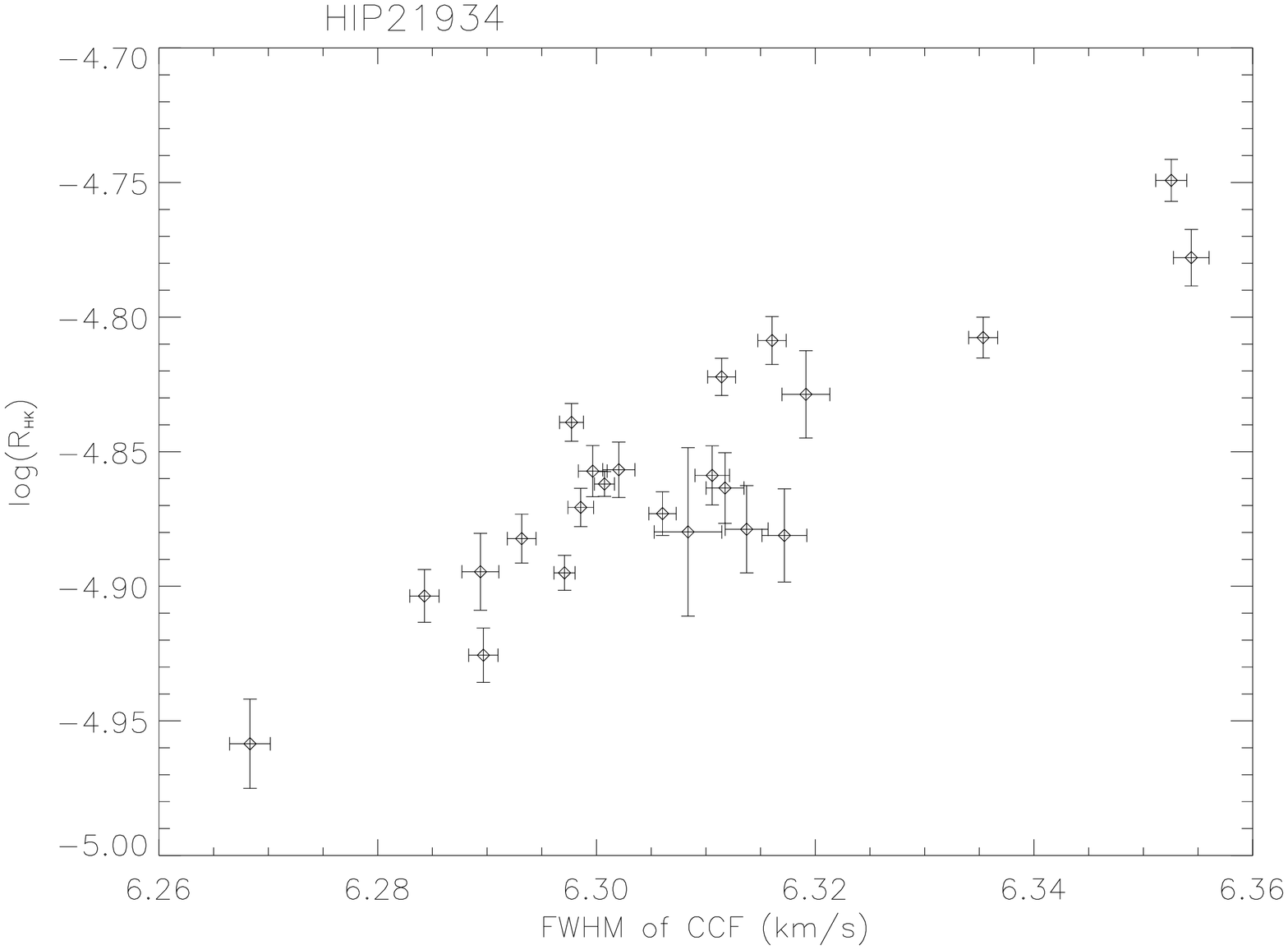,width=6cm,angle=90}
\caption{Variations of the activity tracer, $logR'_{HK}$ or Mount-Wilson $S$ index, are correlated to the full-width-at-half maximum of the cross-correlation function of stars \b (top) and \j (bottom). This indicates that the long-term velocity variation is probably caused by the magnetic cycle of the star, rather than a long-period planetary companion. The correlation is more significant for star \j.}
\label{cycle}
\end{figure}

Twenty-three spectra of \j were secured with HARPS from January 2004 to April 2010, with an average error of 1.4 m/s. Low-amplitude velocity variations are detected, with clearly a long period variation and an additional short-period component (Figure \ref{timecy}). However, as for \b, the full-width-at-half maximum of the cross-correlation function is correlated to the activity index, with a coefficient of 89\% (Figure \ref{cycle}). This again indicates that the velocity variation may be caused by pure stellar activity fluctuations, with a slowly evolving distortion of stellar lines over several years. In this case, the correlation is stronger, and the interpretation in terms of magnetic cycle is favoured, although the star is moderately active during most of the cycle. The velocity variation of this cycle mimics the signature of a long-period (1100 days) planetary companion of minimum mass less than one Jupiter mass.  This result is worse than expected in dedicated studies by \citet{santos2010a}, because the amplitude of the effect is higher - even though \j is a K dwarf, as was considered in this study. It illustrates the importance of monitoring and considering activity tracers simultaneously to radial velocities for a valid interpretation of the results of planet search surveys, especially when low-velocity amplitudes and long orbital periods are considered. The length of the magnetic cycle, if confirmed as such, is 3-3.5 years, again a rather short period compared with the sample of \citet{baliunas95}. This would correspond to a $\sim$ 7-yr magnetic cycle, intermediate between the 2-yr cycle of $\tau$ÊBoo \citep{fares09} and 22-yr cycle of the Sun.

%A fit of an orbital solution with a single companion derives a 1147-day eccentric orbit of a 0.6 \mjup planet. The residuals to this solution are relatively large (1$\sigma\, rms=$4.8 m/s) compared to the individual errors. 
%The best-fit Keplerian solution is found with 0.8 \mjup planet at 1240 day orbital period, plus an inner low-mass planet of 11 day period and 0.1 \mjup mass. Residuals to the fit have a lower scatter of 2 m/s, but given the low number of measurements and the shallow amplitude of the shorter-period component of the system, we favor  in this work the first solution with a single planet. Again, more data are required to complete the analysis, to confirm or deny the presence of a second companion, with a sampling adapted to both $\sim$ 10 day and 1100 day period orbits.

\section{Conclusion}
\label{ccl}

\begin{figure}
\epsfig{file=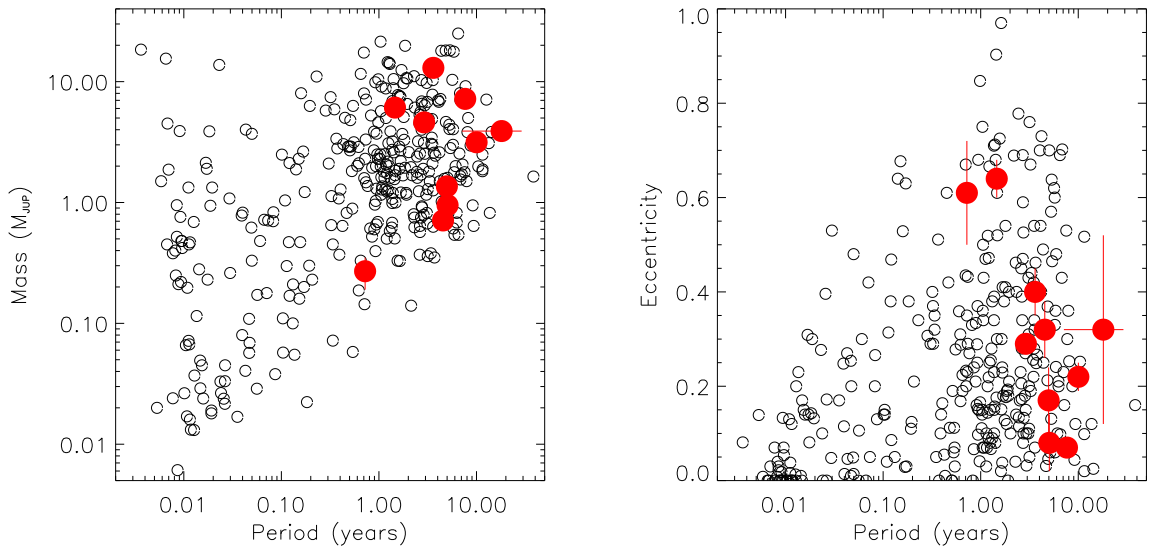,angle=90,width=16cm}
\caption{Mass-period and period-eccentricity diagrams. Open circles show the full sample of radial-velocity planets, as of July 2010, and the filled circled depict the planets described here. Green filled circles related by a line show the location of planet \d b considering both solutions that are equivalent (see text); only one of those dots is a planet.}
\label{stat}
\end{figure}

We presented the observational material secured with the high-precision radial-velocity spectrograph HARPS on ten stars in search for the signature of planetary systems. One star, \e,  had a published planet, for which we could refine the mass determination and orbital elements. Figure \ref{stat} shows how these new planets hold on the total distribution of mass, period, and eccentricity of known exoplanets.

After the first seven years of our survey, we aim at giving first synthetic views of the results by publishing our current detections of long period companions in the planetary mass regime. The radial-velocities time series reported in this work sometimes appear incomplete, and several different solutions cannot be distinguished, as illustrated for some cases by our $\chi^2$ distributions against related parameters $P$, $K$ and $e$. Measurements are continuing however, and in a few years we will be able to better constrain the orbits and refine mass determination of the depicted systems. The discovered companions are giant planets and one or two brown dwarfs, with orbital periods more than three years. In two cases, indications of velocity variations caused by stellar long-term activity evolution are shown, rather than planetary signatures.\\

%\begin{itemize}
Finding long-period, massive planets in volume-limited surveys is of prime importance to get the complete picture of extrasolar system's architecture, despite the natural biases towards short-period planets of the radial-velocity method. However, when searching for best-fit solutions of planetary orbits with periods larger than the time span of the observations, one meets the degeneracy of several types of solutions, with or without linear drifts, additional components in the system, or accounting for degeneracies in the fitted parameters. One expects long period planets to be part of systems, because inward migration should not have blown out the inner planets, and series of giant planets may exist as in the solar system. Orbital fitting for long-period planets is thus more uncertain than for shorter-period planets.  The significance of periods, masses, and eccentricities of known radial-velocity long-distance planets should then be taken with some care, as the best-fit solution is likely to evolve when more data are available. One recent example of revisiting radial-velocity solutions of long-period planets of 47 UMa is illustrated by \citet{gregory2010}. The solutions given in the present work will also be more constrained with additional data, either to complete a period, or to confirm the presence of an additional component in the system. The data are still being collected with HARPS under the observing programme 085.C-0019 (P.I. Lo Curto) and will be presented and analysed in forthcoming papers.\\

Direct imaging of the new planetary systems whose discovery is reported here will be possible, especially for the most nearby target \g. The angular separation presently expected for their planets is 0.152 arcsec from its stars. %With a high minimum mass of 4.6 \mjup, and thus a relatively favorable contrast for direct imaging, the nearby M0 star \k would be the most promising target to be observed with future generation high-dynamics imaging systems such as VLT/SPHERE \citep{beuzit}.  Note that this is an already-known stellar system, containing a distant M3 companion \citep{HIP1993}.\\

As for other planets around giant stars,  \a b is a massive gaseous planet, similar to the recently discovered HD\,32518 b \citep{doellinger2009}. There are now about 20 known planetary systems to giant stars, all with semi-major axes exceeding 0.59 AU and minimum masses larger than 1.5 \mjup. This narrower spread of parameters of planets around giant stars than around main-sequence stars may be explained by the high residual velocity caused by radial oscillations. With periods of hours to days and amplitude of several tens of m/s \citep{hatzes2007}, these intrinsic variations of the star increase the difficulty of detecting the low-mass end of the planets. Regarding arguments such as the filling of the Roche lobe by the primary, the minimum period expected for a giant of 1.09Ê\msol\, mass as \a is about 200 days \citep{mermilliod1996}, which is below the observed 532 day orbit.\\

It may be a new fact that magnetic cycles of stars may have a signature on the radial-velocity behaviour over long timescales, with amplitudes similar to the gravitational perturbation of a giant planet in a large-distance orbit. This was recently discarded by \citet{santos2010a} for a limited sample of G and early K stars, and is further discussed by Lovis et al. (in prep). Two examples are given here, of stars from the volume-limited sample followed-up by HARPS (\j and \b) showing the width of stellar lines affected by the long-term magnetic cycle rather than affected by a distant planetary companion.  In this paper, we tentatively interpret the radial-velocity variations as caused by this long-term cycle, but investigations should be continued in this field, from analyses of a larger sample and from simulations.

One planetary system is presented around \g  with a Saturne-Jupiter-like pair of planets at 0.7 and 2.4 AU respectively. Like the vast majority of planet pairs in systems, the most massive planet is the outer planet. This is possibly a detection bias towards the long periods, because the signal amplitude greatly decreases with star-planet distance.\\

Interestingly, most of the planet host stars discussed here
are metal-poor, contradicting the usual conclusion that stars
with giant planets are usually more metal-rich than the average field
dwarfs (e.g. \citet{santos2004}, \citet{fischer2005}. In a recent
paper, \citet{santos2010b} have also discovered three giant planets
orbiting metal-poor stars from the HARPS survey, all in long period
orbits.  Similarly, it has been noted that the metallicity distribution of giant stars hosting planets does not favour metal-rich stars \citep{pasquini07}. Though preliminary, these facts suggest that long period giant
planets are not uncommon around low-metallicity stars, as shown by a dozen other examples (see http://exoplanet.eu/catalog.php). This possible trend will be confirmed by this continuing survey and others.\\

Finally, let us stress that discoveries of giant planets in very long orbits is only beginning, because of the long baselines that are needed. When the star activity is not an issue, Jupiter twins will be soon collected by radial-velocity surveys, and such systems will be good candidates when searching for Earth twins. The role of Jupiter-like planets as shields for the formation of inner terrestrial planets has been underlined decades ago (e.g. \citet{wetherill1991}). Finding the most massive planet in a system, on a distant orbit, may thus be a first step to discovering planetary systems similar to the solar system, and searching for Earth analogs.

\begin{table*}
\caption{Observed and inferred stellar parameters for the planet-hosting stars presented here. For stars with lowest masses, the relationships which derive the activity index and projected velocity are not calibrated, and the flux in the blue calcium lines is too faint for an estimate of $\log R'_{\mathrm{HK}}$. }
\label{TableStars}
\centering
\begin{tabular}{l l c c c c c c}
\hline\hline
\multicolumn{2}{l}{\bf Parameter} & \bf \a & \bf \b & \bf \c & \bf \d & \bf \e & \bf \f \\
\hline 
Sp & & K1III & K7V & F8V & G3V & G2V & F7V  \\
$V$ & [mag] & 9.17 & 10.02 & 7.79 & 6.71 & 7.46 & 7.64 \\
$B-V$ & [mag] & 1.354 & 1.263 & 0.554 & 0.662 & 0.612 & 0.523\\
$\pi$ & [mas] & 3.22 (1.43) & 36.65 (1.73) & 18.19 (0.45) & 18.69 (0.49) & 20.07 (0.75)& 17.65 (0.55)\\
$d$ & [pc] &  310 (100) & 27.3  (1.3) & 55 (1.5) & 53.5 (1.5) &   49.8 (2) & 57 (2)\\
$M_V$ & [mag] & 1.7 (0.7)& 7.84 (0.1) & 4.09 (0.07) & 3.07 (0.06) & 3.97 (0.1) & 3.86 (0.07) \\
$B.C.$ & [mag] &  -0.75& -0.61 &-0.03 &-0.09 & -0.06& -0.015\\ 
$L$ & [$L_{\odot}$] & 33.1 & 0.1&1.89  &5.11 &2.17&2.30 \\
$T_{\mathrm{eff}}$ & [K]   &4393 (85)  & 4594 (134) & 6160 (65)  & 5740 (62) & 5926 (62) & 6314 (65) \\
log $g$            & [cgs] & 2.12 (0.17) & 4.34 (0.31) & 4.43 (0.1) & 3.97 (0.1) & 4.24 (0.1)  & 4.49 (0.12) \\
$\mathrm{[Fe/H]}$  & [dex] &-0.32 (0.06) & -0.47 (0.08) & -0.11(0.044) & -0.13 (0.04) & -0.02 (0.098)& +0.04 (0.043)\\
$M_*$ & [M$_{\odot}$]  & 1.09 (0.15)&  0.7 (0.1) & 1.09 (0.03) &1.23 (0.03)  & 1.06 (0.04) & 1.21 (0.03) \\
$v\sin{i}$ & [km s$^{-1}$] & 3.5 & - & 1.0 & 2.6 & 1.9 & 4.1\\
$S_{MW}$ & & 0.11 (0.04) & 0.75 (0.06) & 0.15 (0.01)& 0.15 (0.01)& 0.16 (0.01)& 0.15 (0.015) \\
$\log R'_{\mathrm{HK}}$ & - & -&- &-4.99  &-5.12  & -4.97 & -4.97 \\
%$P_{\mathrm{rot}}$ & [days] & &  &  &  & & \\
Age &[Gy]                   & 6.7 (3.2) & 4. (3.7)   & 4.0 (1.6)  & 4.4 (0.3) & 6.5 (1.5) & 1.2 (0.9) \\
$R_*$ & [R$_{\odot}$]  & 16.7 (3.6)& 0.7  & 1.18 (0.04) & 2.23 (0.09)  & 1.36 (0.06) & 1.25 (0.04) \\
\hline
\end{tabular}
\label{stars1}
\end{table*}

\begin{table*}
\caption{Observed and inferred stellar parameters for the planet-hosting stars presented in this paper. (cont'd)}
\label{TableStars2}
\centering
\begin{tabular}{l l c c c c c }
\hline\hline
\multicolumn{2}{l}{\bf Parameter} & \bf \g & \bf \h & \bf \i & \bf \j \\
\hline 
Sp &  & K9V &  G5V &  F8V & K3  \\
$V$ & [mag] & 9.02& 7.90 & 7.80& 9.9  \\
$B-V$ & [mag] &  1.362 & 0.705&0.578  & 1.197 \\
$\pi$ & [mas] &   63.03 (1.36) &  17.29 (1.04) & 18.23 (0.72) & 28.77 (0.97) \\
$d$ & [pc] &  15.8 (0.4) & 58 (3) & 54.8 (2) & 34.8 (1.2) \\
$M_V$ & [mag] & 8.03 (0.06) & 4.08 (0.1) & 4.11 (0.07) & 7.19 (0.08)\\
$B.C.$ & [mag] & -0.66& -0.135 &-0.05  & -0.62 \\ 
$L$ & [$L_{\odot}$] & 0.09& 2.09& 1.89 &0.19\\
$T_{\mathrm{eff}}$ & [K]   & 4685 (155) & 5474 (65) & 5966 (65)  & 4674 (172)\\
log $g$            & [cgs] &  4.28 (0.37)& 4.43 (0.1) & 4.35 (0.11)  & 4.2 (0.41)\\
$\mathrm{[Fe/H]}$  & [dex] & -0.17 (0.074)& -0.29 (0.043) & -0.135 (0.043) &$+$0.025 (0.12) \\
$M_*$ & [M$_{\odot}$]  & 0.7 (0.1) &0.96 (0.02) & 1.02 (0.03) &0.72 (0.02) \\
$v\sin{i}$ & [km s$^{-1}$] &- & $<$ 1&1.4  & - \\
$S_{MW}$ & & 1.05 (0.08)& 0.17 (0.015)& 0.15 (0.01)& 0.57 (0.06) \\
$\log R'_{\mathrm{HK}}$ &- & -&-4.99& -4.86 & -4.86 \\
%$P_{\mathrm{rot}}$ & [days] & &  &  &   & \\
Age &[Gy]                   & 1.3 (1.3) & 11 (1)  & 7.6 &3.9 (3.8)  \\
$R_*$ & [R$_{\odot}$]  & 0.65 (0.1) & 1.7 (0.1) & 1.27 (0.06)   & 0.7 (0.1) \\
\hline
\end{tabular}
\label{stars2}
\end{table*}

\begin{table*} 
\caption{Orbital and physical parameters for the planets presented in
this paper. $T$ is the epoch of periastron. $\sigma$(O-C) is the residual
noise after orbital fitting of the combined set of measurements. $linear$ is the slope of the observed drift with time. }
\label{TablePlanets}
\centering
\begin{tabular}{l l c c c c c }
\hline\hline
\multicolumn{2}{l}{\bf Parameter}&
& \bf \a\,b & \bf \c\,b & \bf \e\,b & \bf \i\,b    \\
\hline
$P$ & [days] & 							&533 (1.7)&1845 (167)&3658 (32)&1319 (4) \\
$T$ & [JD-2400000] & 					&54449 (5)&55301 (449)&55351 (59)&56053 (7)  \\
  $e$ &            &  						&0.64 (0.04)&0.08 (0.06)&0.22 (0.03)&0.40 (0.05) \\
  $linear$& [m/s/yr] &                     			&-7.2 (0.4)&-&-&- \\
$\gamma$ & [km s$^{-1}$] &   				&18.23 (0.03)&43.6 (0.01)&-18.15 (0.001)&9.98 (0.006) \\ 
$\omega$ & [deg]    & 					&122 (8)&96 (89)&-99.3 (6.5)&100 (2)\\
$K$ & [m s$^{-1}$] &   					&190 (29)&15.0 (3.6)&41.0 (1.3)&261 (20) \\
$a_1 \sin{i}$ & [10$^{-3}$ AU] & 			&7.1 (1)&2.5 (0.3)&13.4 (0.4)&28.8 (1.5)\\ 
$f(m)$ & [10$^{-9} $M$_{\odot}$] & 			&181 (84)&0.66 (1.2)&24.2 (2.2)&1857 (307) \\
$m_2 \sin{i}$ & [M$_{\mathrm{Jup}}$] & 		&6.1 (0.9)&0.95 (0.10)&3.15 (0.14)&13.0 (0.8)  \\
$a$ & [AU] &   							&1.3 (0.02)&3.02 (0.16)&4.74 (0.08)&2.38 (0.04) \\
\hline					
$N_{\mathrm{meas}}$ & &				&42&24	&50$a$&15 \\
$Span$ & [days]       &  					&2547&2314	&4515&1538 \\
$\sigma$ (O-C) & [m s$^{-1}$] & 			&29&3.4	&4.2&2.5 \\
$\chi^2_{red}$& & 						&1.6&1.9	&3.7&2.1 \\
\hline
\label{sol1}
\end{tabular}
\end{table*}

\begin{table*} 
\caption{Orbital and physical parameters for \g b and c$^{1}$.}
\label{TablePlanets}
\centering
\begin{tabular}{l l c c}
\hline\hline
\multicolumn{2}{l}{\bf Parameter}
& \bf \g\,b & \bf \g\,c  \\
\hline
$P$ & [days] & 263.3 (2.3) & 1657 (48) \\
$T$ & [JD-2400000] & 54950 (8.5) & 56110 (515)\\
  $e$ &            & 0.61 (0.11) &0.32 (0.06) \\
$\omega$ & [deg]    & 74 (20) & -90 (21) \\
$K$ & [m s$^{-1}$] & 15.5 (12)& 16.5 (1.4)  \\
$a_1 \sin{i}$ & [10$^{-3}$ AU] & 0.26 (0.07)&  2.36 (0.15) \\
$f(m)$ & [10$^{-9} $M$_{\odot}$] &0.048 (0.10)&   0.65 (0.14) \\
$m_2 \sin{i}$ & [M$_{\mathrm{Jup}}$] &0.27 (0.08)&0.71 (0.06) \\
$a$ & [AU] & 0.71 (0.01) &2.43 (0.06) \\
\hline
$\gamma$ & [km s$^{-1}$] & 39.23 (0.012)  \\
$N_{\mathrm{meas}}$ &&29\\
$Span$ & [days]       & 2268 \\
$\sigma$ (O-C) & [m s$^{-1}$] & 1.8 \\
$\chi^2_{red}$ & &1.5\\
\hline
\label{solg}
\footnotetext[1]{Note that there is a marginal possibility for the radial velocity series of this star to be caused by activity rather than planetary signatures (see text).}
\end{tabular}
\end{table*}

\begin{table*} 
\caption{Orbital and physical parameters for the companions with incomplete orbits.}
\label{TablePlanets}
\centering
\begin{tabular}{l l c c c}
\hline\hline
\multicolumn{2}{l}{\bf Parameter}
& \bf \d\,b & \bf \f\,b  & \bf \h\,b\\
\hline
$P$ & [days] &					2798 (60) &6601(-3570,+4141)& 1814 (100)\\
$T$ & [JD-2400000] &  			55053 (93)&60067 (-4954,+4425))&55656 (123)\\
 $e$ &            & 					0.07 (0.01)&0.32 (0.2)&0.17 (0.07)\\
 $linear$& [m/s/yr] &               		-78.9 (0.07) & - & - \\
$\gamma$ & [km s$^{-1}$] &  		44.2 (0.001)&31.7 (0.002)&-57.24 (0.01)   \\
$\omega$ & [deg]    & 			141 (15)&-22 (144)&-80 (24) \\
$K$ & [m s$^{-1}$] & 			91 (1.1)&45.5 (2.3)&23.7 (3)\\
$a_1 \sin{i}$ & [10$^{-3}$ AU] & 	23.4 (0.7)&22 (17)&3.9 (0.4)  \\
$f(m)$ & [10$^{-9} $M$_{\odot}$] &	218 (12)&38 (10)&2.4 (2.3)\\
$m_2 \sin{i}$ & [M$_{\mathrm{Jup}}$] & 7.2 (0.3) &3.9 (0.4)&1.36 (0.12)\\
$a$ & [AU] & 					   4.17 (0.09)&6.8 (-2.4,+3.3)&2.9 (0.12)\\
\hline
$N_{\mathrm{meas}}$ && 26&39&26 \\
$Span$ & [days]       &2465 &2297&1977 \\
$\sigma$ (O-C) & [m s$^{-1}$] &2.4&3.0&2.6\\\
$\chi^2_{red}$ &&3.5&1.9&3.1\\
\hline
\label{sol2}
\end{tabular}
\end{table*}

\begin{acknowledgements}
NCS would like to thank the support by the European Research Council/European Community under the FP7 through a Starting Grant, as well from Fundacao para a Ci\^encia e a Tecnologia (FCT), Portugal, through a contract funded by FCT/MCTES (Portugal) and POPH/FSE (EC), and in the form of grants reference PTDC/CTE-AST/098528/2008 and PTDC/CTE-AST/098604/2008. We are grateful to the ESO staff for their support during observations. Thanks are owing to the anonymous referee for her/his good comments and suggestions to improve the paper.
\end{acknowledgements}

\bibliographystyle{aa}
\bibliography{references}

\onecolumn

\end{document}